\documentclass{emulateapj}
\usepackage{graphicx}
\usepackage{apjfonts}

\newcommand{\kmsend}{\mbox{km s$^{-1}$}} \newcommand{\kms}{\mbox{km
s$^{-1}$ }} 

\newcommand{\hkpc}{\mbox{h$^{-1}$kpc }}

\newcommand{\hmsunend}{\mbox{h$^{-1}$M$_{\sun}$}}
\newcommand{\msun}{\mbox{M$_{\sun}$ }}
\newcommand{\msunend}{\mbox{M$_{\sun}$}}

\newcommand{\lsun}{\mbox{L$_{\sun}$ }}
\newcommand{\lsunend}{\mbox{L$_{\sun}$}}

\newcommand{\cmthree}{\mbox{cm$^{-3}$ }}

\newcommand{\msunyr}{\mbox{M$_{\sun}$yr$^{-1}$ }}
\newcommand{\msunyrend}{\mbox{M$_{\sun}$yr$^{-1}$}}

\newcommand{\htwo}{\mbox{H$_2$}}
\newcommand{\z}{\mbox{$z$}}

\newcommand{\zga}{\mbox{$z \ga$}}
\newcommand{\zla}{\mbox{$z \la$}}
\newcommand{\zsim}{\mbox{$z\sim$ }}
\newcommand{\sdss}{\mbox{J1148+5251 }}
\newcommand{\sdssend}{\mbox{J1148+5251}}

\newcommand{\magorrian}{\mbox{$M_{\rm BH}$-$M_{\rm bulge}$ }}
\newcommand{\Lbol}{L_{\rm {bol}}}
\newcommand{\MBH}{M_{\rm{BH}}}

\newcommand{\Msun}{M_{\odot}}
\newcommand{\Lsun}{L_{\odot}}

\newcommand{\tQ}{\tau_{QSO}}

\newcommand{\Mvir}{M_{\rm{vir}}}
\newcommand{\Vvir}{V_{\rm{vir}}}
\shorttitle{The Nature of CO Emission in \z \ $\sim$ 6 Quasars}
\shortauthors{Narayanan et al.}

\slugcomment{Accepted by ApJ, 15, July, 2007}

\begin{document}
\title{The Nature of CO Emission from \z \ $\sim$
6 Quasars}  
\author{Desika Narayanan\altaffilmark{1}, 
Yuexing Li\altaffilmark{2}, 
Thomas J. Cox\altaffilmark{2}, 
Lars Hernquist\altaffilmark{2}, 
Philip Hopkins\altaffilmark{2},
Sukanya Chakrabarti\altaffilmark{2,3}, 
Romeel Dav\'e\altaffilmark{1},
Tiziana Di Matteo\altaffilmark{6}, 
Liang Gao\altaffilmark{7},
Craig Kulesa\altaffilmark{1}, 
Brant Robertson\altaffilmark{4,5},
Christopher K. Walker\altaffilmark{1}}

\altaffiltext{1}{Steward Observatory, University of Arizona, 933 N
Cherry Ave, Tucson, AZ, 85721, USA}

\altaffiltext{2}{Harvard-Smithsonian Center for Astrophysics, 60
Garden Street, Cambridge, MA 02138, USA} \altaffiltext{3}{NSF
Postdoctoral Fellow} 

\altaffiltext{4}{Kavli Institute
for Cosmological Physics and Department of Astronomy and Astrophysics,
University of Chicago, 933 East 56th St., Chicago, Il,, 60637}

\altaffiltext{5}{Spitzer Fellow} 

\altaffiltext{6}{Carnegie Mellon
University, Department of Physics, 5000 Forbes Ave., Pittsburgh, PA
15213} 

\altaffiltext{7} {Institute for Computational Cosmology,
Dept. of Physics, University of Durham, South Road, Durham, DH1 3LE,
UK} 

\begin{abstract}
We investigate the nature of molecular gas emission from \zsim 6
quasars via the commonly observed tracer of \htwo, carbon monoxide
(CO). We achieve these means by combining non-local thermodynamic
equilibrium (LTE) radiative transfer calculations with merger-driven
models of \zsim 6 quasar formation that arise naturally in
$\Lambda$-cold dark matter ($\Lambda$CDM) structure formation
simulations. Motivated by observational constraints, we consider four
representative \zsim 6 quasars formed in the halo mass range $\sim
10^{12} - 10^{13}$ \msun from different merging histories.  Our main
results are as follows. We find that, owing to massive starbursts and
funneling of dense gas into the nuclear regions of merging galaxies,
the CO is highly excited during both the hierarchical buildup of the
host galaxy and the quasar phase and the CO flux
density peaks between J=5-8. The CO morphology of \zsim 6 quasars
often exhibits multiple CO emission peaks which arise from molecular
gas concentrations which have not yet fully coalesced. Both of these
results are found to be consistent with the sole CO detection at \zsim
6, in quasar \sdssend.  Quasars which form at \zsim 6 display a large
range of sightline-dependent line widths. The sightline dependencies
are such that the narrowest line widths are when the rotating
molecular gas associated with the quasar is viewed face-on (when the
$L_B$ is largest), and broadest when the quasar is seen edge on (and
the $L_B$ is lowest). Thus, we find that for all models selection
effects exist such that quasars selected for optical luminosity are
preferentially seen to be face-on which may result in CO detections of
optically luminous quasars at \zsim 6 having line widths narrower than
the median. The mean sightline-averaged line width is found to be
reflective of the circular velocity of the host halo, and thus scales
with halo mass. For example, the mean line width for the $\sim 10^{12}$
\msun halo is $\sigma \sim 300$ \kmsend, while the median for the
$\sim 10^{13}$ \msun quasar host is $\sigma \sim 650$ \kmsend.
Depending on the host halo mass, approximately 2-10\% of sightlines in
our modeled quasars are found to have narrow line widths compatible
with observations of \sdssend. When considering the aforementioned
selection effects, these percentages increase to 10-25\% for quasars
selected for optical luminosity.  When accounting for both temporal
evolution of CO line widths in galaxies, as well as the redshift
evolution of halo circular velocities, these models can
self-consistently account for the observed line widths of both
submillimeter galaxies and quasars at \zsim 2. Finally, we find that
the dynamical mass derived from the {\it mean} sightline-averaged line
widths provide a good estimate of the total mass, and allows for a
massive molecular reservoir, supermassive black hole, and stellar
bulge, consistent with the local \magorrian relation.

\end{abstract}

\section{Introduction}
\label{section:introduction}

The discovery of extremely luminous quasars at $z\ga$6 via novel color
selection techniques demonstrates that massive galaxies and 
supermassive black
holes formed very early in the Universe (Fan et al. 2002, 2003, 2004).
Observations of the dusty and molecular interstellar medium (ISM) in
\zsim 6 galaxies can serve as a unique probe into the star formation
process in the first collapsed objects, and help quantify the
relationship between star formation and black hole growth when
the Universe was less than a billion years old
(e.g. Wang et al. 2007; for a recent review, see
Solomon \& Vanden Bout 2005). Serving as a proxy for observationally
elusive molecular hydrogen (\htwo), rotational transitions in tracer
molecules such as $^{12}$CO (hereafter, CO), HCN and HCO$^+$ can
provide diagnostics for the physical conditions in the star-forming
giant molecular clouds (GMCs) of high redshift galaxies
(e.g. Bertoldi et al. 2003a,b; Carilli et al. 2005; Riechers et
al. 2006a; Walter et al. 2003).

The highest redshift quasar that has been found, SDSS J1148+5251
(hereafter \sdssend) at \z=6.42 (Fan et al. 2003) is an extremely
bright object with bolometric luminosity $\sim 10^{14}$ \lsunend,
and is thought to be powered by accretion onto a supermassive black
hole of mass $\sim 10^9$ \msun (Willott, McLure, \& Jarvis 2003).
Bertoldi et al. (2003b) measured a far infrared luminosity of
$1.3 \times 10^{13}$ \lsunend, which, if powered solely by
starburst-heated dust, corresponds to an exceptional
star formation rate of
$\sim 3000$ \msunyrend.

Pioneering CO observations of \sdss have revealed a great deal
concerning the molecular ISM in the host galaxy of this \zsim 6
quasar. Through multi-line observations and large velocity gradient
(LVG) radiative transfer modeling, Bertoldi et al. (2003a) found that
the CO flux density peaks at the J=6 level of CO, indicative of the
warm and dense conditions characteristic of vigorous star formation.
A measured molecular gas mass of $M_{\rm H_2} \approx 10^{10}$ \msun
shows that \sdss plays host to a large reservoir of molecular gas
(Bertoldi et al. 2003a; Walter et al. 2003). Subsequent high
resolution observations with the Very Large Array (VLA) by Walter et
al. (2004) discovered that the CO emission in this galaxy is extended
on scales of 2.5 kpc and is resolvable into two emission peaks separated
by 1.7 kpc, with each peak tracing $\sim 5 \times 10^{9}$ \msun of
molecular gas. These observations suggest that \sdss is a merger
product (Walter et al. 2004; Solomon \& Vanden Bout 2005).

CO observations of \sdss have presented challenges for models
of galaxy formation as well.  For example, dynamical mass estimates
from observed CO line widths are not able to account for the presence
of a $\sim 10^{12}$ \msun stellar bulge as would be predicted by the
present-day \magorrian relation, suggesting that the central
supermassive black hole could have grown in part before the host
galaxy (Walter et al. 2004). In contrast, the presence of heavy
elements (Barth et al. 2003), and significant CO emission (Bertoldi et
al. 2003a; Walter et al. 2003, 2004) imply that the ISM has been
significantly enriched with metals from early and abundant star
formation. Recent theoretical arguments have additionally proposed
that a relation between black hole mass and stellar bulge mass is a
natural consequence of AGN feedback in galaxies (Di Matteo et al.
2005, 2007; Hopkins et al. 2007a; 
Sijacki et al. 2007), and that this relation shows only
weak ($\sim 0.3-0.5$ dex) evolution in galaxies from redshifts \z=0-6
(Robertson et al. 2006a,c; Hopkins et al. 2007a).

Numerical simulations can offer complementary information to
observations of \zsim 6 quasars by providing a framework for the
formation and evolution of \zsim 6 quasars and the relationship
between the star forming ISM in these galaxies and observed CO
emission. Calculations by Springel, Di Matteo \& Hernquist (2005a,b)
have found that galaxy mergers serve as a viable precursor for quasar
formation. Strong gaseous inflows driven by tidal torques on the gas
(e.g. Barnes \& Hernquist 1991, 1996) can fuel nuclear starbursts
(e.g. Mihos \& Hernquist 1994, 1996) and feed the growth of central,
supermassive black holes (Di Matteo et al. 2005);
subsequent feedback from the AGN can lift
the veil of obscuring gas and dust and, along numerous sight lines,
reveal an optically bright quasar (Hopkins et al. 2005a,b).

More recently, Li et al. (2007) have proposed a merger driven model
for quasar formation at \zsim 6 which fits naturally into a
$\Lambda$CDM framework. By performing numerical simulations which
simultaneously account for black hole growth, star formation, quasar
activity, and host spheroid formation, these authors found that galaxy
mergers in early $\sim 10^{13}$ \msun halos can result in the
formation of bright quasars at \zsim 6.  The quasars in these
simulations exhibit many properties similar to the most luminous
quasars at \zsim 6, including both observables, such as the rest-frame
$B$-band luminosity, and inferred characteristics (e.g.  the central
black hole mass and bolometric luminosity). 
These simulations can thus serve as a laboratory for investigating the
properties of the interstellar medium in the first quasars, as well as
the relation between CO emission and star formation at early epochs.

In order to quantitatively couple these models with CO observations,
molecular line radiative transfer calculations are necessary. Recent
works by Narayanan et al. (2006a,b) have developed a methodology for
simulating molecular line transfer on galaxy-wide scales.  Here, we
aim to investigate the plausibility of \zsim 6 quasar formation in
massive $\sim 10^{12} - 10^{13}$ \msun halos in a hierarchical
structure formation scenario by modeling the observed CO emission from
these high redshift sources. We achieve these means by coupling the
non-LTE radiative transfer codes of Narayanan et al. (2006a,b) with
the hierarchical \zsim 6 quasar formation models of Li et al.
(2007). We make quantitative predictions for \zsim 6 quasars which
form in halos with 
masses ranging from $\sim 10^{12} - 10^{13}$ \msunend,
and provide interpretation for existing and future CO observations of
$z\ga$6 quasars. 

In \S~\ref{section:numericalmethods} we describe the numerical models
employed. In \S~\ref{section:excitation}, we present the modeled CO
excitation characteristics and luminosities. In
\S~\ref{section:morphology}, we discuss the morphology of the CO gas.
In \S~\ref{section:lines} we describe the derived emission line
profiles and compare our models to observations of the only detected
CO measurement in a \zsim 6 quasar, \sdssend. In
\S~\ref{section:mdyn}, we use the model CO lines to investigate the
usage of CO observations as dynamical mass indicators in the highest
redshift quasars. We conclude with a discussion comparing the CO
emission properties of our simulated quasar with other high-\z\ galaxy
populations in \S\ref{section:discussion} and summarize in
\S~\ref{section:conclusions}. Throughout this paper we assume a
cosmology with $h$=0.7, $\Omega_\Lambda$=0.7, $\Omega_M$=0.3.

\section{Numerical Methods}
\label{section:numericalmethods}

 In order to capture the physics of early Universe quasar formation,
simulations must have the dynamic range to faithfully track the
evolution of the most massive halos within which these 
quasars reside (e.g. Haiman \& Loeb 2001), as well as follow the
stars, dark matter, ISM and black holes in the progenitors of the
quasar host galaxy. We have performed multi-scale calculations which
include cosmological dark matter simulations in a volume of 3
Gpc$^3$ to identify the most massive halos, and subsequent hydrodynamic
galaxy merger computations within these halos at higher resolution. The
cosmological and galaxy merger simulations were implemented using the
parallel, $N$-body/Smoothed Particle Hydrodynamics (SPH) code GADGET-2
(Springel 2005). We then applied the non-LTE radiative transfer code,
{\it Turtlebeach} (Narayanan et al. 2006a,b), to the outputs of the
hydrodynamic galaxy merger simulations to investigate the
emission properties of the CO molecular gas in the resultant
quasar. The quasar formation simulations and radiative transfer
algorithms are described in detail in Li et al. (2007) and Narayanan
et al.  (2006a,b), respectively, and we refer the reader to those
works for more detail; here, we briefly summarize, and focus on the
aspects of the calculations directly relevant to this study.

\subsection{Cosmological Simulations}

Observational evidence suggests that quasars at \zsim 6 likely form in
very massive ($\sim 10^{13}$ \msunend) halos. For example, quasars at
\zsim 6 have an extremely low comoving space density of $n
\approx 10^{-9}$ Mpc$^{-3}$ (e.g. Fan et al. 2003), comparable to
the rarity of massive $\sim 10^{13}$ \msun halos at \zsim 6.  Further
evidence comes from black hole mass-halo mass correlations, and
theoretical arguments relating quasar luminosity and halo mass (Lidz
et al. 2006). In addition, recent numerical simulations by Pelupessy, Di
Matteo \& Ciardi (2007)
have suggested that the buildup of supermassive
$\sim 10^{9}$ \msun black holes by \zsim 6 may require galaxy mergers
in extremely rare $\gtrsim 10^{13}$ \msun halos. The strongest
evidence for high redshift quasar formation in massive halos comes
from recent SDSS clustering measurements by Shen et al. (2007) which
indicate that the minimum mass for high redshift quasar hosts is
$\sim 6-9 \times 10^{12}$ \msunend. While these arguments are
suggestive, they are not conclusive. Thus, one major goal of this work
is to further investigate the hypothesis that \zsim 6 quasars live in
high mass halos and constrain the range of host halo masses
for these sources by comparing their simulated CO emission to
observations. To this end, we have simulated 4 quasars which
form hierarchically through numerous mergers in massive halos
ranging in mass from $\sim 10^{12} - 10^{13}$ \msunend.

 In order to simulate the large subvolume of the Universe necessary to
track the evolution of massive halos and formation of \zsim 6 quasars,
as well as achieve a reasonable mass resolution at the same time, we
performed the structure formation simulation with a multi-grid
procedure similar to Gao et al. (2005). We first ran a uniform
resolution simulation of a 3 Gpc$^3$ volume, with an effective mass
and spatial resolution of $m_{\rm dm} \sim 1.3 \times 10^{12}$ \msun
and $\epsilon\sim 125$ \hkpc (comoving softening length),
respectively, with initial conditions generated by CMBFAST (Seljak \&
Zaldarriaga 1996). We assumed a $\sigma_{\rm 8}$ of 0.9 (though see
Li et al. [2007] for a discussion of the implications of other choices
for the cosmological parameters).

We then used a friends-of-friends group finder in order to seek out
candidate massive halos within which early quasars form. Anticipating
that these rare quasars will be progenitors of massive objects today,
we identified the most massive halos at \z=0, 
and then resimulated the evolution of these objects and their immediate
environment at a much higher mass and force resolution assuming an
initial redshift of \z=69. The final effective resolutions of the halo
evolution simulations were $m_{\rm{dm}}\sim 2.8 \times 10^8$ \msun
and $\epsilon\sim5$ \hkpc.

The merger tree of the halos was extracted from the simulations in
order to provide information concerning the masses of the largest
contributing progenitors to the halo mass and allow us to reconstruct
the hierarchical buildup of the quasar host galaxy. We considered
groups that contribute at least 10\% of the halo mass as progenitors
in the merger history resulting in e.g. seven progenitors for the most
massive halo simulated.  Table~\ref{table:ICs} lists the properties of
all the resultant quasars.

The discussion throughout this paper will largely focus on quasars
Q1-Q3 which formed hierarchically in the cosmological
simulations. Quasar Q4 is a binary coplanar merger specially simulated
for studying the dependence of the CO emission properties on merger
history, and will be discussed only in \S~\ref{section:mergerhistory}.

\begin{deluxetable*}{lcccccc}
\tabletypesize{\scriptsize}

\tablecaption{Quasar Models\label{table:ICs}}
\tablewidth{0pt}
\tablehead{
\colhead{Name\tablenotemark{a}} & 
\colhead{\z\tablenotemark{b}}  &
\colhead{$\Mvir$\tablenotemark{c}} & 
\colhead{$\Vvir$\tablenotemark{d}} &
\colhead{$\MBH$\tablenotemark{e}} & 
\colhead{$\Lbol(\rm {peak})$\tablenotemark{f}} &
\colhead{$\tQ(\Lbol \ge 10^{13}\, \Lsun)$\tablenotemark{g}}\\
&
&
\colhead{[$10^{12}\, \Msun$]} &
\colhead{[km s$^{-1}$]}           &
\colhead{[$10^{9}\, \Msun$]}&
\colhead{[$10^{13}\, \Lsun$]}&
\colhead{[Myr]}
}
\startdata
Q1 & 6.5 &   7.7 & 626.4 & 2.0 & 2.0 & 55.5  \\
 Q2 & 5.6 &   3.8 & 467.9 & 1.0 & 5.0 & 22.3  \\
 Q3 & 7.2 &   1.5 & 340.7 & 0.2 & 1.5 & 4.0   \\
 Q4 & 5.5 &   8.1 & 593.2 & 0.4 & 1.1 & 21.9  \\
\enddata

\tablenotetext{a}{Name of quasar model.}
\tablenotetext{b}{Redshift of the quasar at peak accretion activity.}
\tablenotetext{c}{Virial mass of the quasar host halo, assuming overdensity $\Delta=200$.}
\tablenotetext{d}{Virial velocity of the quasar host halo, assuming overdensity $\Delta=200$.}
\tablenotetext{e}{Black hole mass of the quasar.}
\tablenotetext{f}{Peak bolometric luminosity of the quasar.}
\tablenotetext{g}{Quasar lifetime for $\Lbol \ge 10^{13}\, \Lsun$.}
\end{deluxetable*}

\subsection{Galaxy Merger Simulations}

To derive the physical properties of the \zsim6 quasar host galaxies,
we resimulated their merger trees hydrodynamically using
GADGET-2. GADGET-2 utilizes a fully conservative SPH formulation which
allows for an accurate handling of discontinuities (Springel \&
Hernquist 2002). The code accounts for radiative cooling of the gas
(Katz et al. 1996; Dav\'e et al. 1999), and a multiphase description
of the ISM that includes cold clouds in pressure equilibrium with a
hot, diffuse gas (e.g. McKee \& Ostriker 1977; see also Springel \&
Hernquist 2003a).  Star formation is constrained by observations of
local galaxies, and follows the Kennicutt-Schmidt laws (Kennicutt,
1998; Schmidt 1959; Springel \& Hernquist 2003b). The progenitor
galaxies had dark matter halos initialized to follow a Hernquist
(1990) profile, and the virial properties are scaled to be appropriate
for cosmological redshifts (Robertson et al. 2006a).

 Black holes in the simulations are realized through sink particles
that accrete gas following a Bondi-Hoyle-Lyttleton parameterization
(Bondi \& Hoyle 1944; Bondi 1952; Hoyle \& Lyttleton 1939). To
model feedback from central black holes, we assume that 0.5\% of the
accreted mass energy is reinjected into the ISM as thermal energy (Di
Matteo et al. 2005; Springel, Di Matteo \& Hernquist 2005a,b).  This
formulation for AGN feedback in galactic scale simulations has been
shown to successfully reproduce X-ray emission patterns in galaxy
mergers (Cox et al. 2006a), observed quasar luminosity functions and
lifetimes (Hopkins et al. 2005a-d, 2006a,c,d, 2007b), the Seyfert galaxy
luminosity function (Hopkins \& Hernquist 2006), the M$_{\rm
BH}$-$\sigma_v$ relation (Di Matteo et al. 2005; Hopkins et al. 2007a;
Robertson et al. 2006a), the bimodal galaxy color distribution
(Springel et al. 2005b, Hopkins et al. 2006b,e, 2007c), characteristic CO
emission patterns in ultraluminous infrared galaxies (ULIRGs;
Narayanan et al. 2006a), infrared colors of ULIRGs and \zsim 2
Submillimeter Galaxies with embedded AGN (Chakrabarti et al. 2007a,b),
and the kinematic structure of merger remnants (Cox et al. 2006b).

We assume that the black holes in these simulations formed from the
first stars (e.g. Abel, Bryan \& Norman 2002; Bromm \& Larson 2004;
Yoshida et al. 2006), and that the seed masses for the black holes are
200 \msun at \z=30. Before the first progenitor galaxies entered the
merger tree, their black holes grew at the Eddington limit, resulting
in seed black hole masses of $\sim 10^{4}$ \msun at the time of the
first merger.  The black holes in the simulations are assumed to merge
when their separation is less than a gravitational softening length
(30 h$^{-1}$pc).

We model evolution of the CO emission in the quasars from $z \approx
$8 to $z \approx $5, noting that the peak of the quasar activity is
roughly 7$\ga z \ga$5.5. The black hole luminosity outshines the
stellar luminosity for a large range of redshifts, and it is this time
that we refer to as the 'quasar phase' for any given galaxy.

\subsection{Evolution of Model Quasars}
 To aid in the discussion in the remainder of this work, we
qualitatively describe the evolution of the most massive
($\sim 10^{13}$ \msunend) quasar host, though the results are general
for all models considered in Table~\ref{table:ICs}. The quasar host
galaxy builds hierarchically, through seven major mergers between z=
14.4-8.5. Strong gravitational torques on the gas drive massive
gaseous inflows, causing heavy accretion onto the central black
hole(s), and triggering intense starbursts that typically form stars
at a rate between $\sim 10^{3} - 10^4$ \msunyrend. The black holes accrete
heavily as gas is funneled in toward the nuclear regions. Feedback
from the most massive central black hole then drives a powerful wind,
creating numerous lines of sight along which the central quasar is no
longer obscured, and the black hole luminosity outshines the stellar
luminosity. The central supermassive black hole can be visible as an
optically bright quasar ($\Lbol > 10^{13}\Lsun$) for $\sim 50$ Myr
(though less for lower mass quasar host galaxies;
Table~\ref{table:ICs}).  The powerful quasar wind quenches the
starburst, and self-regulates the black hole growth. In the post
quasar phase, the luminosity of the galaxy subsides, and it eventually
evolves into a cD-like galaxy. More details concerning the evolution
of the models presented here are discussed in Li et al. (2007).

\subsection{Molecular Line Radiative Transfer}
\label{section:ratran}
The CO emission properties were calculated using {\it Turtlebeach}, a
3-dimensional non-LTE radiative transfer code, based on an expanded
version of the Bernes (1979) algorithm (Narayanan et al. 2006a,b). Our
improvements focus on including a mass spectrum of GMCs in a subgrid
manner, which allows us to more accurately model the molecular line
radiation from the dense cores of molecular clouds, as well as the
diffuse outer layers.

We build the emergent spectrum by integrating the equation of
radiative transfer along various lines of sight:
\begin{equation}
  I_\nu = \sum_{r_0}^{r}S_\nu (r) \left [ 1-e^{-\tau_\nu(r)} \right
  ]e^{-\tau_\nu(\rm tot)}
\end{equation}
where $I_\nu$ is the frequency-dependent intensity, $S_\nu$ is the
source function, $r$ is the physical depth along the line of sight,
and $\tau$ is the optical depth. The source function is dependent on
the CO level populations which are assumed to be in statistical
equilibrium, but not in LTE. This aspect of the calculation is
important as the assumption of LTE often breaks down when considering
the propagation of submillimeter-wave radiation through the molecular
ISM.

 We consider collisions with \htwo, radiative excitation, and
stimulated and spontaneous emission in determining the CO level
populations. The radiative transfer is handled in a Monte Carlo
manner, in which photon `packets' are emitted isotropically over
4$\pi$ steradians. The frequencies are randomly drawn from the local
line profile function which includes effects from the kinetic
temperature and a microturbulent velocity field which we assume to
have a constant value of 0.8 \kmsend. The strongly density-dependent
collisional excitation rates are modeled by deriving the density
distribution in a given cell from a mass spectrum of giant molecular
clouds. In this formulation, we model the density distributions in the
clouds as singular isothermal spheres with power-law index $p$=2
(e.g. Walker, Adams \& Lada 1990), where the radius of each cloud is
determined by the Galactic GMC mass-radius relation (e.g. Solomon et
al. 1987; Rosolowsky 2005, 2006). Our results for higher lying CO
transitions are particularly improved using this subgrid non-LTE
approach. The molecular mass fraction of cold gas is assumed to be
half, as motivated by local volume surveys (e.g. Keres, Yun \& Young,
2003). This resulted in molecular gas masses of
$1-3\times10^{10}$\msunend in the model quasars.

Our methodology is an iterative one. A solution for the CO level
populations is first guessed and provides the initial radiation
field. We then calculate the mean intensity field in a Monte Carlo
manner. The contributions of the radiation field and collisional
excitation rates determine the updated level populations via rate
equations and the process is repeated until convergence.

We have benchmarked our radiative transfer codes against the published
non-LTE line transfer tests of van Zadelhoff et al. (2002), and
present the results in Narayanan et al. (2006b). Our approach in
combining non-LTE radiative transfer calculations with galaxy-scale
hydrodynamic simulations has predicted CO morphologies, spatial
extents, excitation levels, line widths and intensities that are in
good agreement with observations of local starbursts and galaxy
mergers (Narayanan et al. 2006a; D. Narayanan et al, in prep.).

For the models presented here, $1\times 10^7$ model photons were
emitted per iteration. The mass spectrum of GMCs in each sub-grid cell
considered clouds with a lower cutoff at $1 \times 10^4$ \msun and an
upper cutoff at $1 \times 10^{6}$ \msunend, consistent with observed
mass ranges of GMCs (e.g. Blitz et al. 2006). We calculated CO
transitions across 10 levels at a time and assumed a uniform Galactic
CO abundance of CO = $1.5\times 10^{-4} \times$ \htwo \ (Lee, Bettens
\& Herbst 1996). We utilized collisional rate coefficients from the
{\it Leiden Atomic and Molecular Database} (Sch\"{o}ier et al. 2005).

Throughout this work, we make predictions for CO transitions ranging
from J=1-0 through J=10-9. One potential caveat regarding the lower
lying (e.g. J=1-0) transitions is that the low density molecular gas
in \zsim 6 quasars may be in a diffuse intracloud medium as is thought
to be the case for local ULIRGs, rather than bound in GMCs (Downes \&
Solomon 1998). In our subgrid modeling of GMCs, we assume implicitly
that the diffuse molecular gas is bound in the outer envelopes of GMCs
which are characterized as singular isothermal spheres. The assumption
of higher density gas residing in bound cloud cores is likely to be
more robust. Recent studies show similarities between LIR/HCN ratios
in both local GMCs and a large sample of ULIRGs. These similarities
suggest that the high density gas even in the nuclei of galaxy mergers
is locked in cloud cores (Wu et al. 2005), similar to the simulations
presented here. In this sense, our modeled environment of higher
density gas is more reasonable, and predictions concerning
higher-lying transitions of CO (e.g. J$\ga$3) more robust.

\section{Excitation and Luminosity of CO}
\label{section:excitation}

\begin{figure}
\scalebox{0.7}{\rotatebox{90}{{\plotone{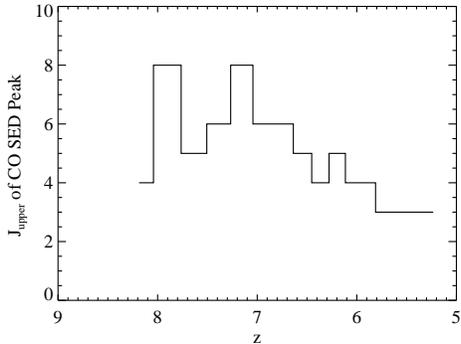}}}}
\caption{Upper rotational state (J$_{\rm upper}$) of CO SED peak as a
function of redshift for most massive quasar host, model Q1 (see
Table~\ref{table:ICs}). During the early, massive starburst, most of
the CO is highly excited. During the height of the quasar phase, the
peak in the CO excitation in our simulated galaxy ranges from J=8 in
the beginning of the quasar phase to J=5 near the end. This is broadly
consistent with observations of excited molecular gas in the sole CO
detection at \zsim 6 (Bertoldi et al. 2003a). As the starburst
subsides in the post-quasar phase, lower temperatures and densities
drive the peak in the CO SED down to
J $\sim 3$.\label{figure:cosed_peak}}
\end{figure}

\begin{figure*}
\scalebox{0.35}{\rotatebox{90}{\plotone{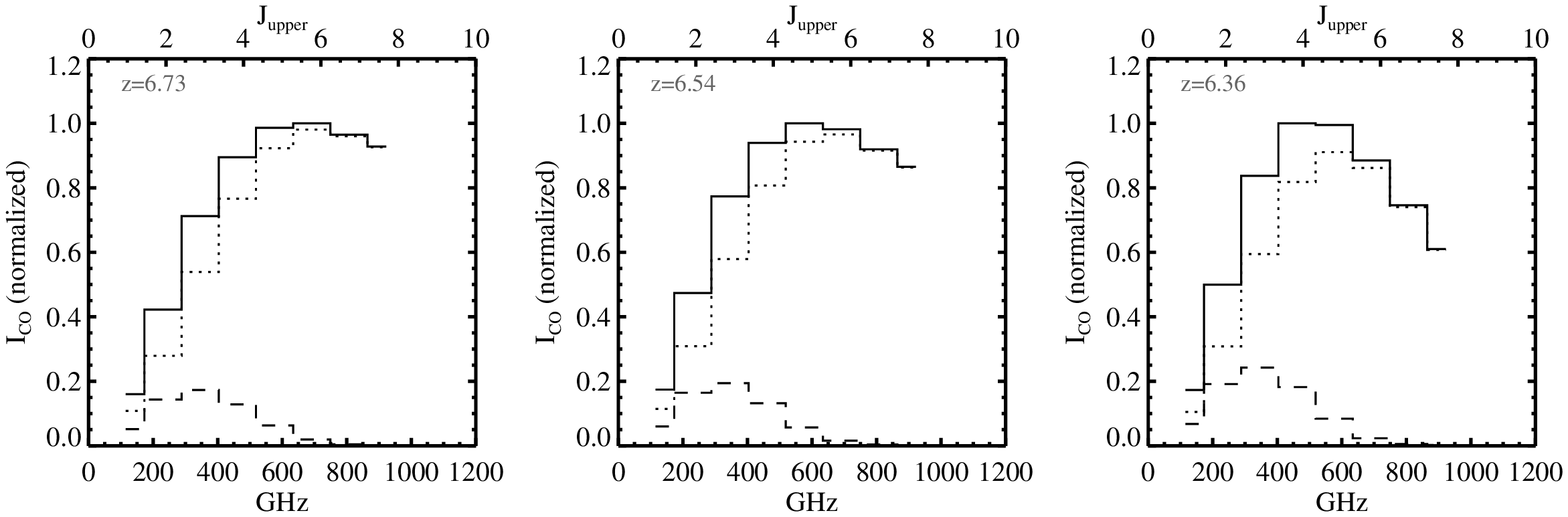}}}
\caption{The CO SED (solid line) at three points during the peak of
the quasar phase of our most massive simulated galaxy (M
$\sim 10^{13}$ \msunend; model Q1). Massive starbursts and dense
conditions in the central kiloparsec cause the CO flux
density to peak at J=8 prior to and at the beginning of the quasar
phase (not shown). During the bulk of the quasar phase, the CO SED
peak drops to J=6 owing to quenching of the starburst by black hole
feedback, and remains roughly constant for the remainder of the
quasar's life. As the central AGN quenches the nuclear starburst, less
gas is in highly excited (J$\ga$6) states, and the slope on the blue
end of the peak becomes much steeper. The dotted line is the CO SED
from just the central kpc, and the dashed line is the contribution
from radius=1-4 kpc). The upper level of each transition is listed on
the top axis.
\label{figure:cosed}}
\end{figure*}

The CO spectral energy distribution (SED) is the CO flux density
emitted per rotational J state and thus provides a direct measure of
the excitation characteristics of molecular gas (e.g. Wei\ss \ et
al. 2005a). The shape and peak of the CO SED describe the relative
number of molecules in a given rotational J state, and serve as
observable diagnostics of the underlying temperature and density of
the emitting molecular gas. Here, we describe in detail the excitation
characteristics of the most massive quasar host (Q1, $M
\sim 10^{13}$ \msunend), and explicitly note when the lower mass
models exhibit different properties. In
Figure~\ref{figure:cosed_peak}, we show the redshift evolution of the
peak in the CO SED. Moreover, in Figure~\ref{figure:cosed}, we
show representative CO SEDs for quasar Q1 at three redshifts during
the quasar phase. The CO SEDs are averaged over three orthogonal
viewing angles, though the SEDs derived from each individual angle are
nearly identical.

Strong gravitational torques exerted on the gas by the multiple
mergers drive massive amounts of cold gas into the central kiloparsec,
giving rise to densities as high as $\sim 10^7$ \cmthree in GMC cores
at the beginning of the quasar phase (\zsim 7). The combination of
these dense conditions and heating associated with the continued
starburst cause the CO molecular gas to become highly excited. During
this time, the peak of the CO SED rises to J=8. Molecular gas in the
central $\sim$ kiloparsec dominates the high-lying excitation. To place
this in the context of active star forming regions in the local
Universe, Wei\ss, Walter \& Scoville (2005b) found the CO SED to peak
at J=6 in the nuclear region of starburst galaxy M82.

 As energy input from the quasar begins to quench the starburst and
dispel gas from the central regions, the star formation rate drops to
$\sim 10^2$ \msunyrend. Consequently, the peak excitation in the
molecular gas rapidly drops to values more similar to local
starbursts. During most of the quasar phase, the CO flux density in
the simulated galaxy peaks at J=6. This is consistent with
the multi-line observations and CO SED derivations by Bertoldi et
al. (2003a), who find the CO flux density in \sdss to peak at J=6. In
a merger-driven model for high-\z \ quasars, the interplay between
massive starbursts and feedback from central black holes is important
in determining the observed CO excitation characteristics in galaxies
like \sdssend.

The peak of the CO flux density remains roughly constant through the
height of the quasar phase. As feedback from the central black holes
further extinguishes the nuclear starburst, fewer molecules are highly
excited, and consequently, the relative flux density from higher J
levels begins to drop. In Figure~\ref{figure:cosed}, the slope of the
CO SED at levels higher than the turnover point is seen to become
steeper as the quasar evolves. When the accretion onto the central
supermassive black hole subsides in the post-quasar phase (\zla 6),
the star formation rate (SFR) drops to $\la$50 \msun yr$^{ -1}$. The
bulk of the molecular gas in this late stage of the galaxy's evolution
is only moderately excited, and the peak in the CO SED declines to
J$\approx$3-4.

The trends discussed above are similar in models Q2 and Q3 (Table
~\ref{table:ICs}), though the overall normalization is slightly
different. The CO SED of both the intermediate mass and low mass
models (Q2 and Q3) peaks at J=6 at the beginning of the quasar phase,
and settles at J=4 as the starburst subsides.
phase. The lower excitation values in the lowest mass model owe to
overall lower densities and star formation rates.
For example, during the quasar phase, the SFR from model Q3 is
$\sim 30$ \msunyrend.

\begin{figure}
\scalebox{.9}{\plotone{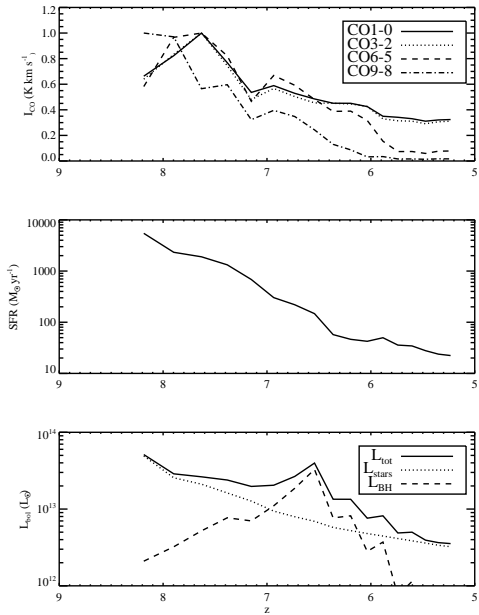}}
\caption{{\it Top}: Velocity-integrated CO Intensity (K \kmsend) as a
  function of redshift for various CO transitions for quasar Q1. Each
transition is normalized. {\it Middle}: Star formation rate and {\it
Bottom} Quasar bolometric luminosity as a function of redshift. The
bolometric luminosity is broken up into contribution by stars and
black holes. The bright CO emission in high J transitions during the
quasar phase is representative of massive star formation, though the
bulk of the spheroid formed during the hierarchical buildup of the
quasar host galaxy, prior to the active quasar phase.
\label{figure:lco}}
\end{figure}

Another way to view the excitation characteristics of the molecular
gas is through the velocity-integrated CO luminosity (in units of
K-\kmsend, where the K is the Rayleigh-Jeans temperature). In
Figure~\ref{figure:lco}, we show the normalized velocity integrated CO
(J=1-0, J=3-2, J=6-5 and J=9-8) intensity as a function of
redshift. We additionally plot the SFR, stellar, and black hole
luminosity. The CO luminosity across all transitions decreases as the
merger activity progresses, and the starburst reaches its peak. This
is especially true of the high-J states which are typically excited by
collisions in the starburst-heated gas.  All CO transitions peak in
integrated intensity early on, when the starburst has not consumed
most of the available star-forming gas, and collisions help to sustain
molecular excitation. As the starburst fades owing to a diminishing
fuel supply, the intensity from the high lying CO transitions (e.g. 6-5,
9-8) falls off rapidly while the lower-J transitions experience a more
moderate decline. In part, this owes to the fact that the lowering of
gas temperatures and densities does not heavily affect the molecular
excitation at J=1. De-excitation of warm, dense star forming gas
additionally contributes to populating the lower J states. While the
CO luminosity is only about half of its maximum value during the
quasar phase, the bolometric luminosity of the galaxy peaks here as
the central quasar becomes visible (Figure~\ref{figure:lco}, {\it
bottom panel}).

\section{Molecular Gas Morphology}
\label{section:morphology}

\begin{figure*}
\scalebox{.6}{\rotatebox{90}{\plotone{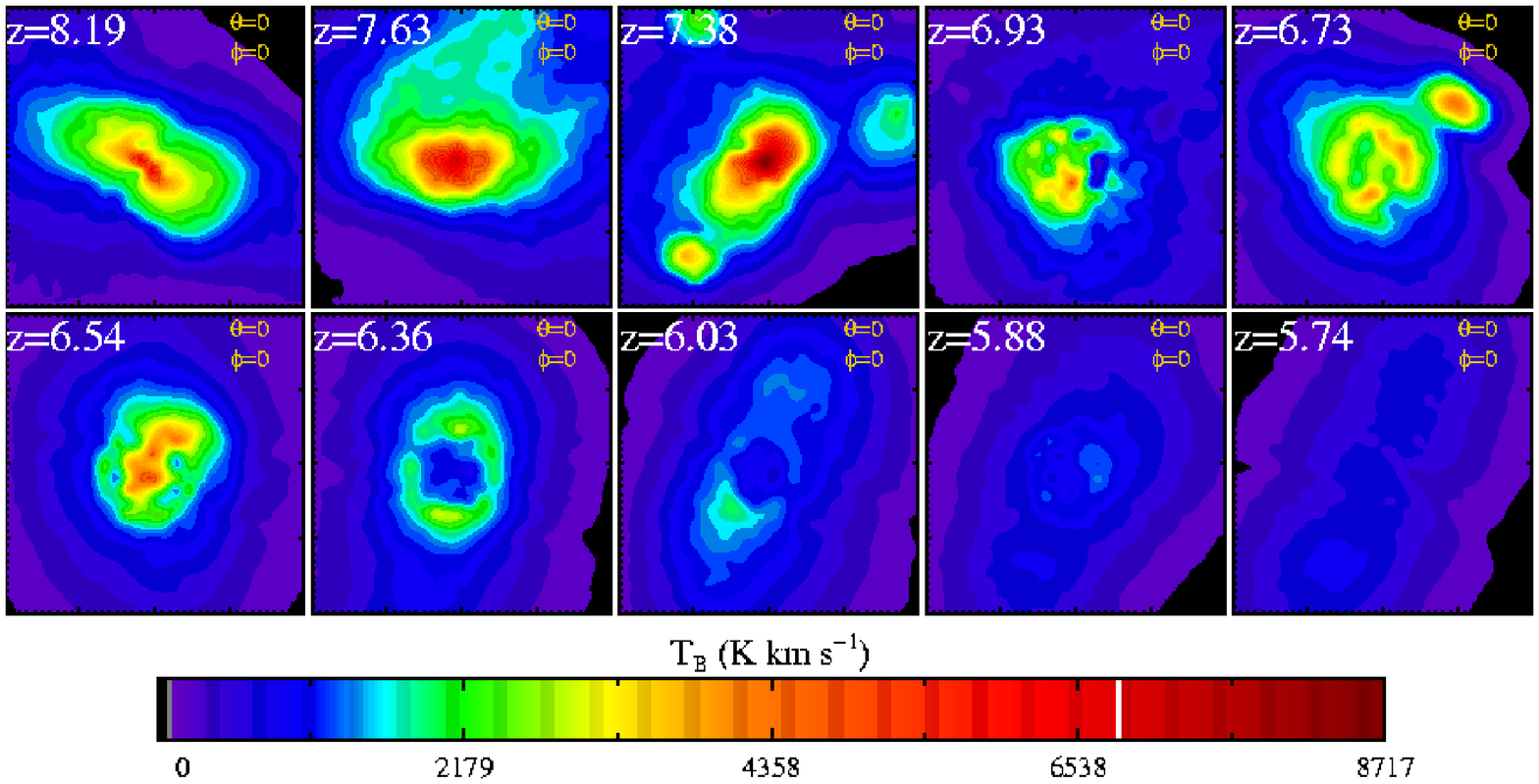}}}
\caption{Evolution of CO (J=1-0) emission contours from \z=8.2-5.7 in
most massive halo, model Q1 ($M \sim 10^{13}$ \msunend). Multiple
emission peaks are visible as cold gas falls in toward the nucleus,
similar to observations of \sdssend.  As the starburst and quasar
activity subsides, the CO intensity fades. Each panel is four kpc on a
side. The emission is in terms of velocity-integrated intensity
(K-\kmsend), and the scale is at the bottom. The viewing angle of each
panel is listed in the top right corner, and is always
$\theta=0$,$\phi=0$ for this figure.\label{figure:mapevolution}}
\end{figure*}

In this section, we discuss the CO morphology of the model quasar host
galaxies through their evolution.  The discussion is again focused on the
most massive simulation (Q1), though the results are generic for each
of the models studied here. In Figure~\ref{figure:mapevolution}, we
show the evolution of the central 2 kpc of the CO (J=1-0) emission in
the most massive halo, Q1, during the hierarchical buildup of the
host galaxy and quasar phase. Individual concentrations in the
molecular gas density which have not fully coalesced appear during the
buildup of the quasar host galaxy, and through parts of the quasar
phase (e.g. \z=6.73, Figure~\ref{figure:mapevolution}), giving rise to
multiple CO (J=1-0) emission peaks.  Near the end of the quasar's
lifetime, the molecular gas settles into a nuclear disk, with the
densest gas in the central $\sim 500$ pc.

\begin{figure*}
\plotone{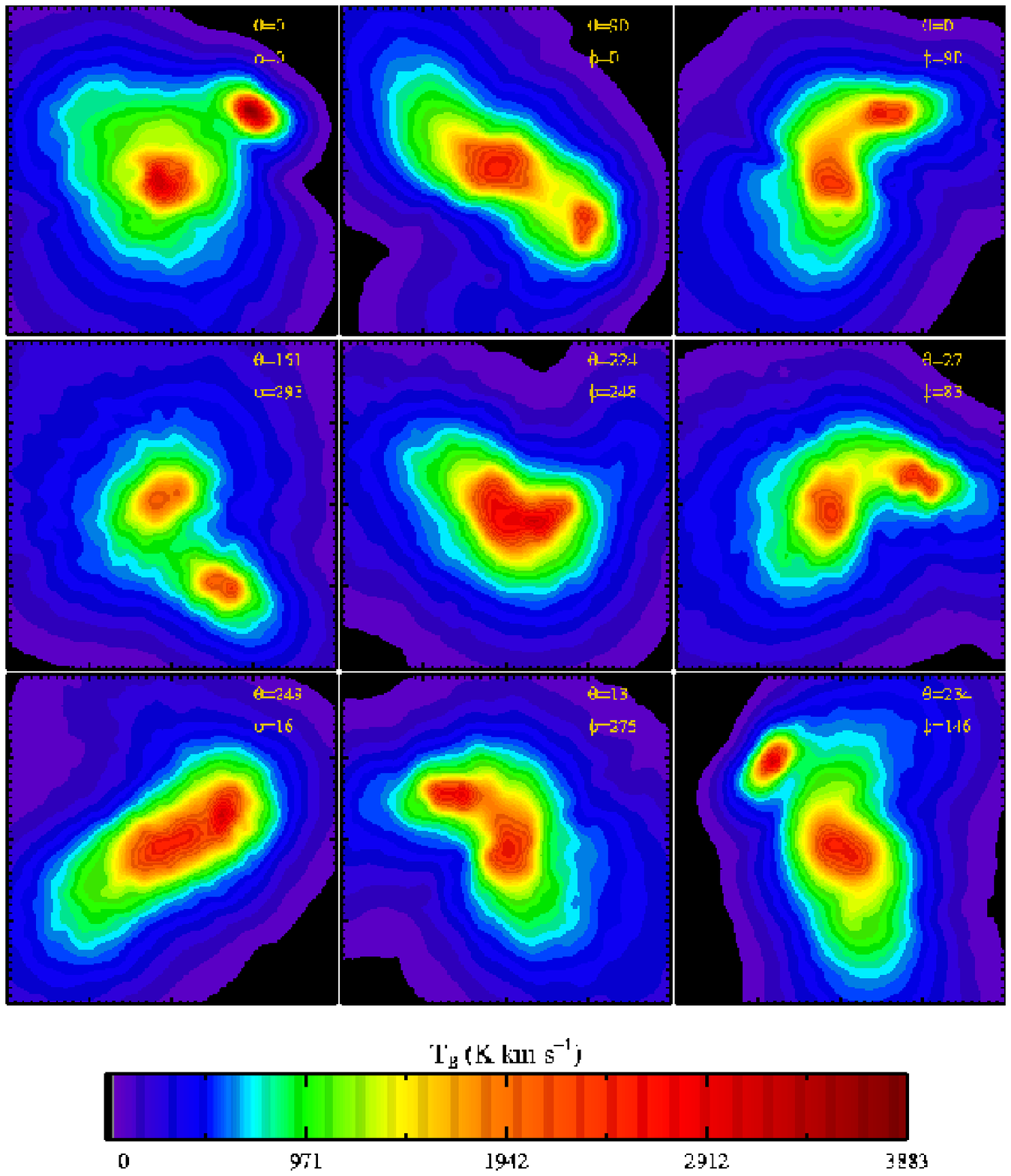}
\caption{CO (J=3-2) emission from most massive quasar host (Q1) at
z=6.73, while cold gas from the merger is still infalling. Top three
panels are three orthogonal viewing angles, and the bottom six viewing
angles are randomly drawn. The orientation for the line of sight is in
the top right corner of each panel. Along many different viewing
angles, multiple CO components are visible, similar to the observed
morphology of \sdss (Walter et al. 2003), indicating a viable merger
origin for \sdssend. Each panel is four kpc on a side.  The emission
is in terms of velocity-integrated intensity (K-\kmsend), and the
scale is at the bottom.  \label{figure:snap8maprandom}}
\end{figure*}

The morphological features of our simulated quasar agree reasonably
well with observations.  Observations of \sdss have revealed two CO
(J=3-2) emission peaks in the central 2 kpc (Walter et al. 2004),
similar to the multiple surface CO surface brightness peaks seen at
many points in our models
(e.g. Figure~\ref{figure:mapevolution}). This suggests that the
observed multiple surface brightness peaks in the CO morphology of
\sdss may owe to separated peaks of high density emission in the
nucleus that have not yet coalesced.  To further illustrate this
point, in Figure~\ref{figure:snap8maprandom}, we have plotted the CO
(J=3-2) emission contours at \z=6.73 over three orthogonal viewing
angles, and six random ones. Most viewing angles exhibit two surface
brightness peaks, suggesting that a merger origin for the formation of
\sdss is viable.

Within the constraints of our numerical simulations, multiple density
peaks in the cold gas appear to be the most plausible explanation for
the observed morphology of \sdssend. Multiple emission peaks in the CO
morphology of galaxy mergers have also been noted to arise from large
entrainments of molecular clouds in AGN-driven winds (Narayanan et
al. 2006a).  In the current simulations, however, the characteristic
outflow morphologies of Narayanan et al. are not seen during the
active quasar phase. Emission from the nuclei of progenitor galaxies
can additionally cause multiple CO surface brightness peaks. However,
by the time the simulated host galaxy reaches the height of the quasar
phase, the most massive nuclei have all merged into the central
potential (Li et al. 2007). 

\begin{figure*}
\epsscale{.9}
\plotone{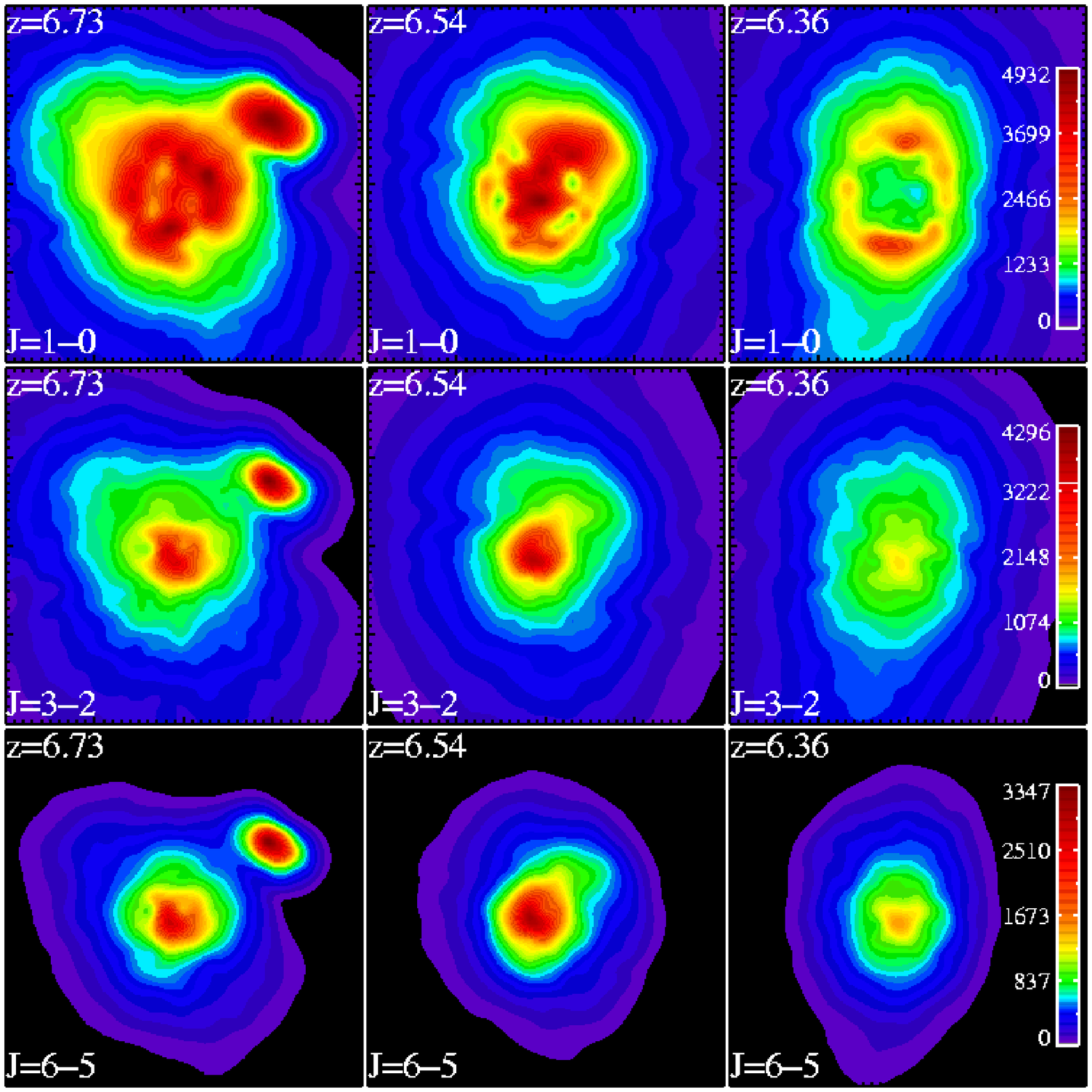}
\caption{Excitation dependent CO morphologies: CO (J=1-0), (J=3-2) and
(J=6-5) emission contours from most massive quasar host (Q1) during the
peak of the quasar phase. Columns are constant in redshift, and rows
are constant in transition. Multiple CO emission peaks are seen
arising from merging cold clumps of gas, as well as circumnuclear gas
in the (J=1-0) case. The gas in the central $\sim$ kpc is highly
excited, and thus does not show prominent (J=1-0) emission, but causes
the (J=3-2) and (J=6-5) emission to be centrally concentrated.
Multiple CO emission peaks are seen in higher lying transitions owing
only to merging gas as the bulk of the gas in the central $\sim$ kpc is
highly excited. These results are robust across all modeled viewing
angles.  The viewing angle for each panel is the same, and is
$\theta=0$, $\phi=0$, for comparison with
Figures~\ref{figure:mapevolution} and ~\ref{figure:snap8maprandom}.
Each panel is four kpc on a side, and each row (transition) is on its
own scale to facilitate interpretation. The scales are on the right,
and the units are in terms of velocity integrated Rayleigh-Jeans
temperature (K-\kmsend).\label{figure:redshift_transition}}
\end{figure*}

The excitation characteristics in the vicinity of massive starbursts
can cause transition-dependent CO morphologies. In
Figure~\ref{figure:redshift_transition}, we show the CO (J=1-0),
(J=3-2), and (J=6-5) emission contours for the central 2 kpc of the
most massive host during the quasar phase. The multiple emission peaks
which owe to merging clumps of dense gas (\z=6.73 in this plot) appear
in all transitions. When the cold gas has coalesced in the nucleus,
emission from the lowest excitation gas traced by CO (J=1-0) exhibits
discrete pockets of emission whereas higher-lying emission originating
from denser gas is centrally concentrated. As seen in
Figure~\ref{figure:cosed}, the warm and dense conditions in the
starbursting nucleus of the quasar ensure that the CO remains excited,
with the peak flux density at J=5-8. The CO gas in the nucleus is
typically excited out of the J=1 level, resulting in discrete pockets
of emission in the circumnuclear molecular gas. The higher-lying
emission, which originates in the warmer, denser cores of GMCs, is
more prevalent in the central 500 pc of the quasar, resulting in a
smoother distribution.

Owing to the violent nature of the quasar's formation process, the
spatial extent of the CO emission has a large dynamic range, and
varies from $\sim 2$ kpc to $\sim 300$ pc (half-light radius) throughout
the evolution of the host galaxy.  The more extended morphologies are
representative of times when gas is falling into the central
potential, and in the post quasar phase when winds have produced an
extended morphology. More compact CO emission is seen primarily when
the cold gas has completely coalesced, and the quasar is most active.

The starburst is centrally concentrated in the central $\sim 250$ pc of
the quasar host galaxy during the majority of the simulation presented
here. Conversely, the CO morphology is not always so compact. The CO
emission through J=10 is extended during the buildup of the host
galaxy (prior to the quasar phase) owing to merging gas clumps.  This
implies that CO may not always serve as an adequate tracer of the
starburst during the major merger phase of the quasar's
formation. During the active quasar phase, when most of the gas has
fallen into the nucleus, emission from higher CO transitions
(e.g. J$\ga$6) tends to become compact, and faithfully trace the
active starburst.

\section{CO Emission Lines}
\label{section:lines}
The simulated properties of the CO emission lines show variations
among the models presented in Table~\ref{table:ICs}. In this section
we discuss the CO emission lines in terms of the CO (J=6-5) lines as
the CO SED peaks near J=6 for most of the quasar phase, and thus this
transition best traces the properties of the bulk of the molecular
gas. Unless otherwise specified, the nature of the CO line profile as
described in the remainder of this section is not seen to vary
significantly with observed transition.

\subsection{General Nature of Modeled Line Profiles}
\label{section:generallines}

\begin{figure*}
\scalebox{.8}{\rotatebox{90}{\plotone{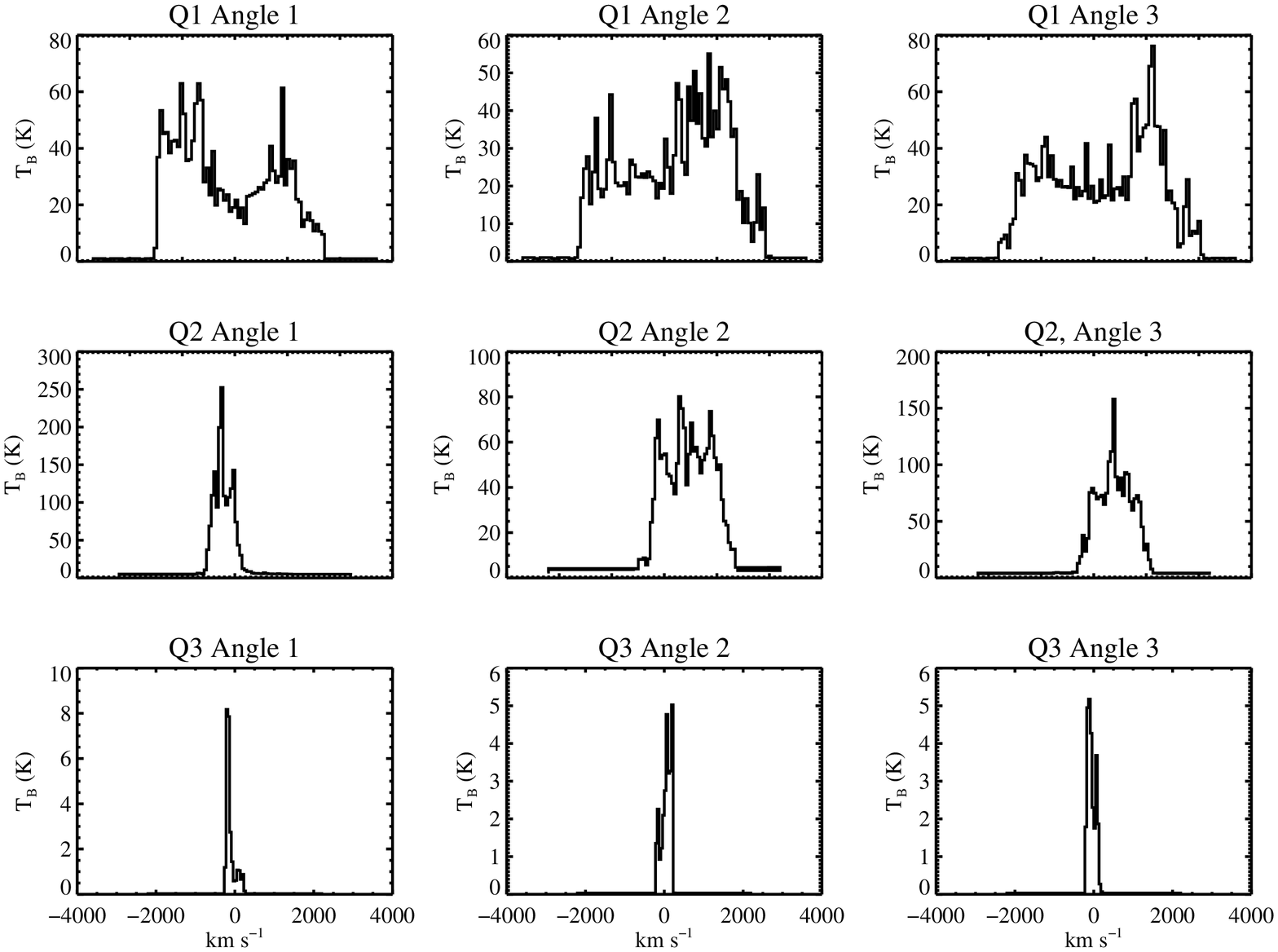}}}
\caption{CO (J=6-5) emission lines from model quasars Q1, Q2 and Q3 at
peak of their quasar phases viewed from three random sightlines.  The
first row corresponds to quasar Q1, the second to Q2 and third row to
Q3, The characteristic emission line widths drop with halo mass,
though remain featured in each model with several substructure spikes
superposed on broader lines. Spectra have been convolved with a
circular $5 \arcsec$ Gaussian beam, and modeled at an angular diameter
distance of 1 Gpc.
\label{figure:snap8specrandom}}
\end{figure*}

In Figure~\ref{figure:snap8specrandom}, we show a sample 3 random CO
(J=6-5) emission lines from the simulated quasars Q1, Q2 and Q3 at the
peak of their respective quasar phases.  The first noteworthy point
regarding the CO emission lines from the \zsim 6 quasar models is that
they are characteristically featured, with a significant amount of
substructure.  Along many lines of sight, several smaller emission
spikes (with $\sigma \sim 50-100$ \kmsend) originating in dense CO
clumps in the central 2 kpc sit superposed on the broader emission
line. These emitting clumps are centrally concentrated in the quasar,
and similar emission features are seen in higher spatial resolution
spectra.

\begin{figure}
\scalebox{0.9}{\rotatebox{90}{\plotone{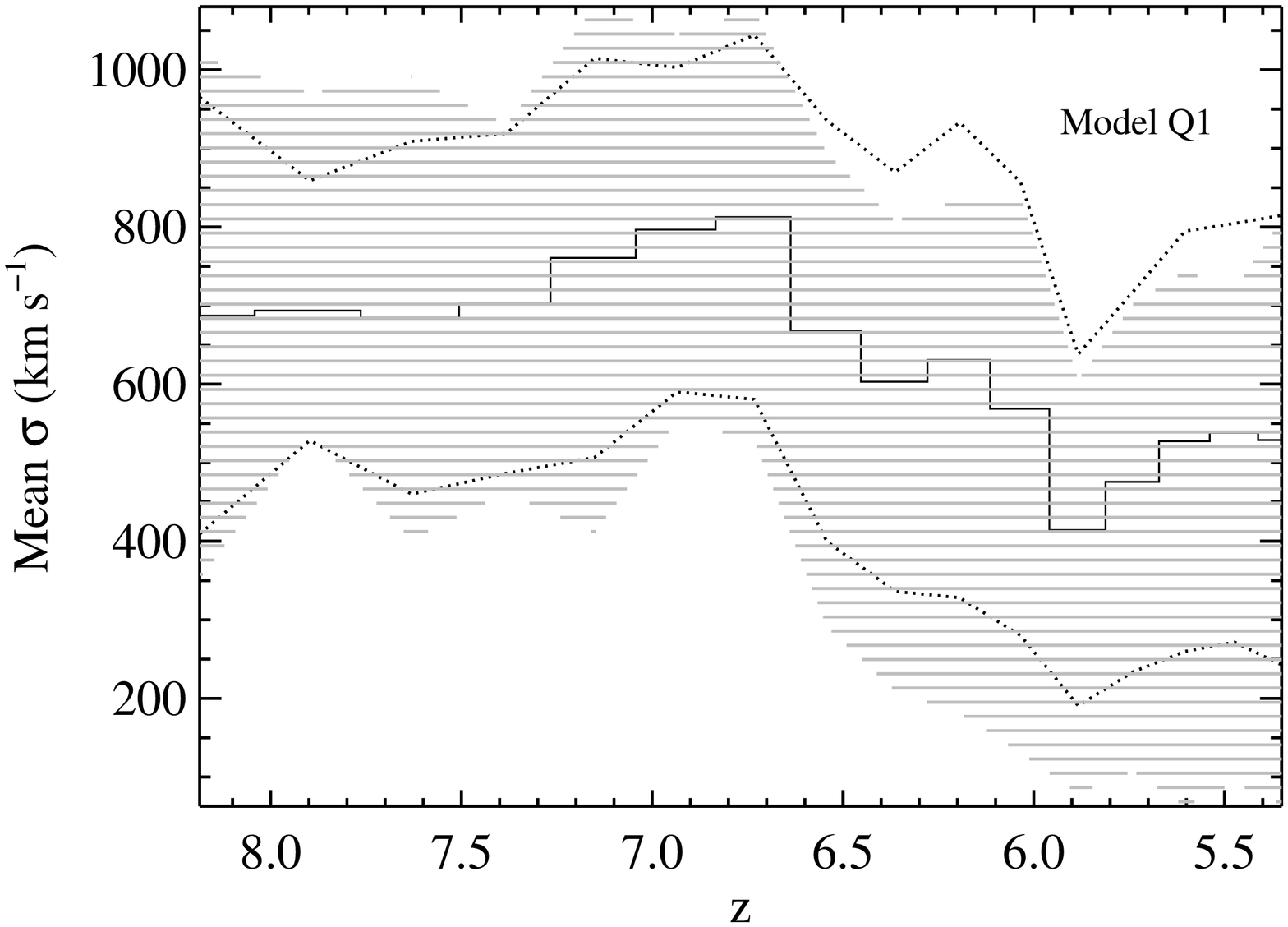}}}
\scalebox{0.9}{\rotatebox{90}{\plotone{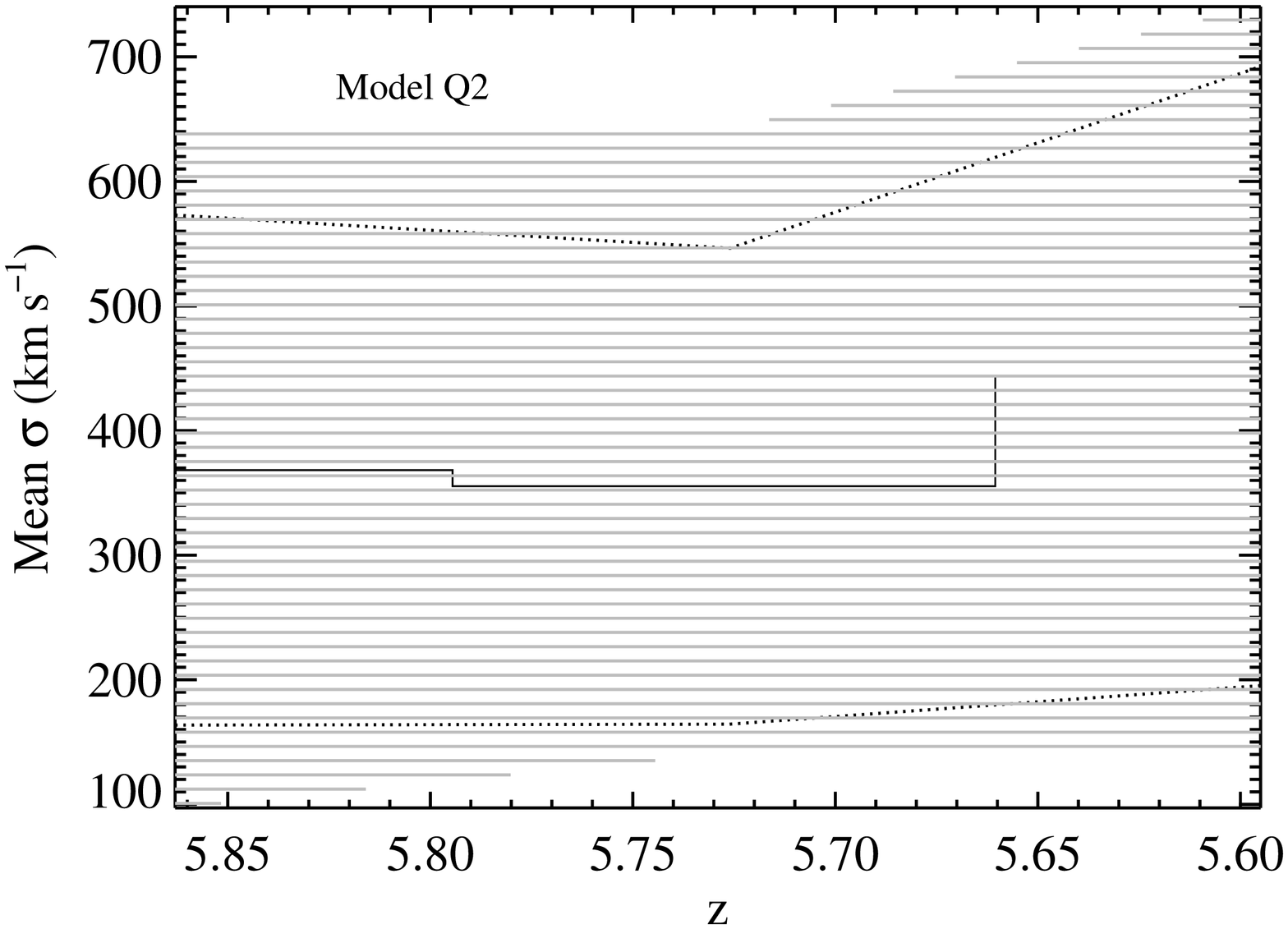}}}
\scalebox{0.9}{\rotatebox{90}{\plotone{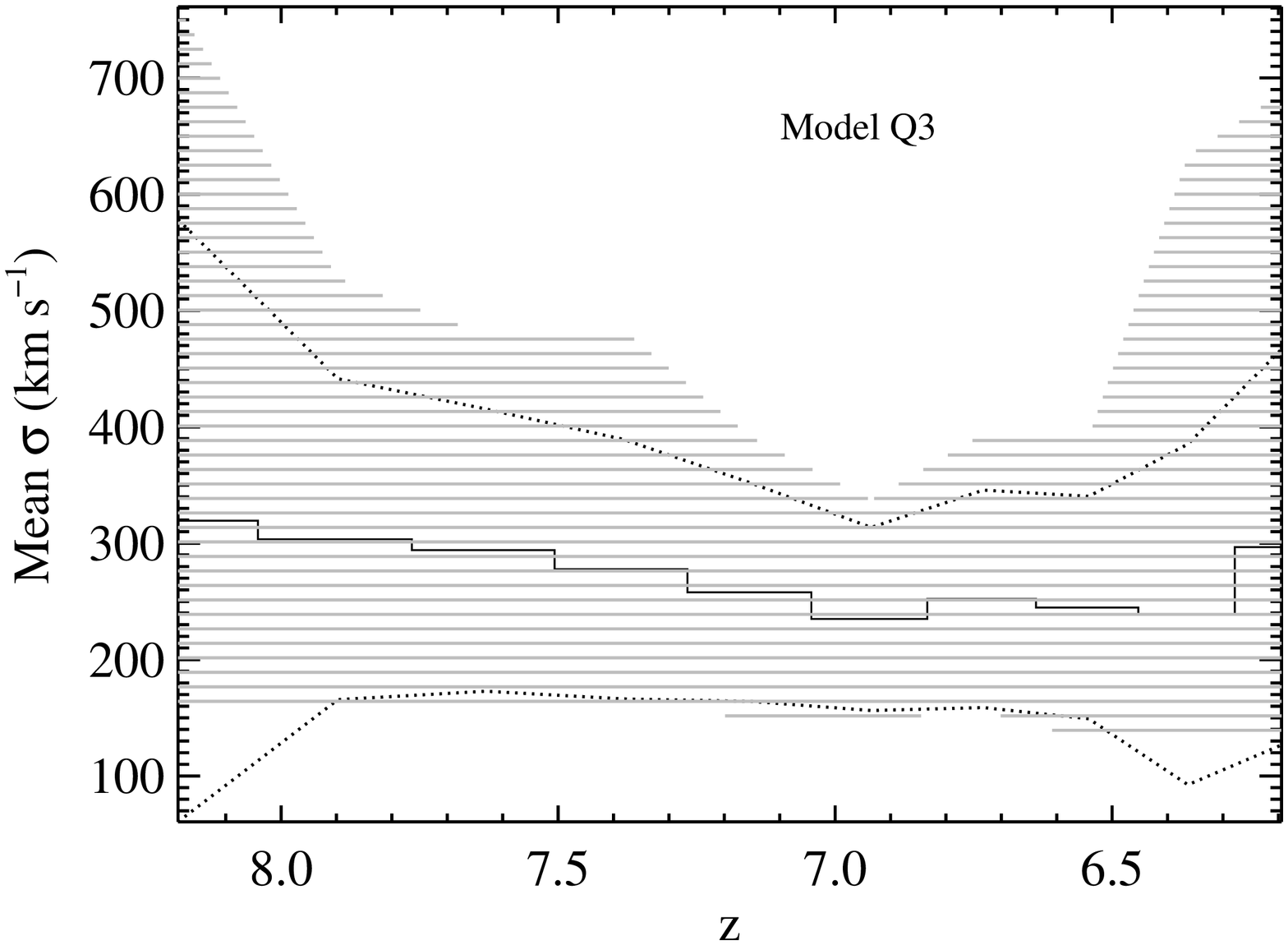}}}
\caption{Sightline-averaged velocity dispersion ($\sigma$) of CO
  (J=6-5) emission lines (solid line, middle) as a function of
  redshift for quasar models Q1-Q3. The shaded area shows the range of
  derived $\sigma$ over 250 randomly sampled sightlines, and the upper
  and lower dotted lines show the $2\sigma$ linewidths for each
  snapshot. The spectra have been convolved with a circular $5 \arcsec$
  Gaussian beam, and binned to 50 \kmsend. During the hierarchical
  buildup of the host galaxy, the \htwo \ gas is highly dynamical, and
  the typical line widths are broader than the virial velocity of the
  host galaxy by a factor of $\sim 1.5-2$. As the gas virializes during
  the quasar phase, the line widths drop, and roughly trace the virial
  velocity of the galaxy. A range of line widths are permitted
  throughout the evolution of the host galaxy, with larger numbers of
  sightlines being compatible with the narrow ($\sigma \sim 120$
  \kmsend) detected line in \sdss near the end of the quasar
  phase. The number of sightlines compatible with observations
  naturally increases in the lower mass halos (Q2 and Q3) as the CO
  lines faithfully trace the virial velocity of the galaxy during the
  quasar phase. \label{figure:meanfwhm}}
\end{figure}

\begin{figure}

\scalebox{0.9}{\rotatebox{90}{\plotone{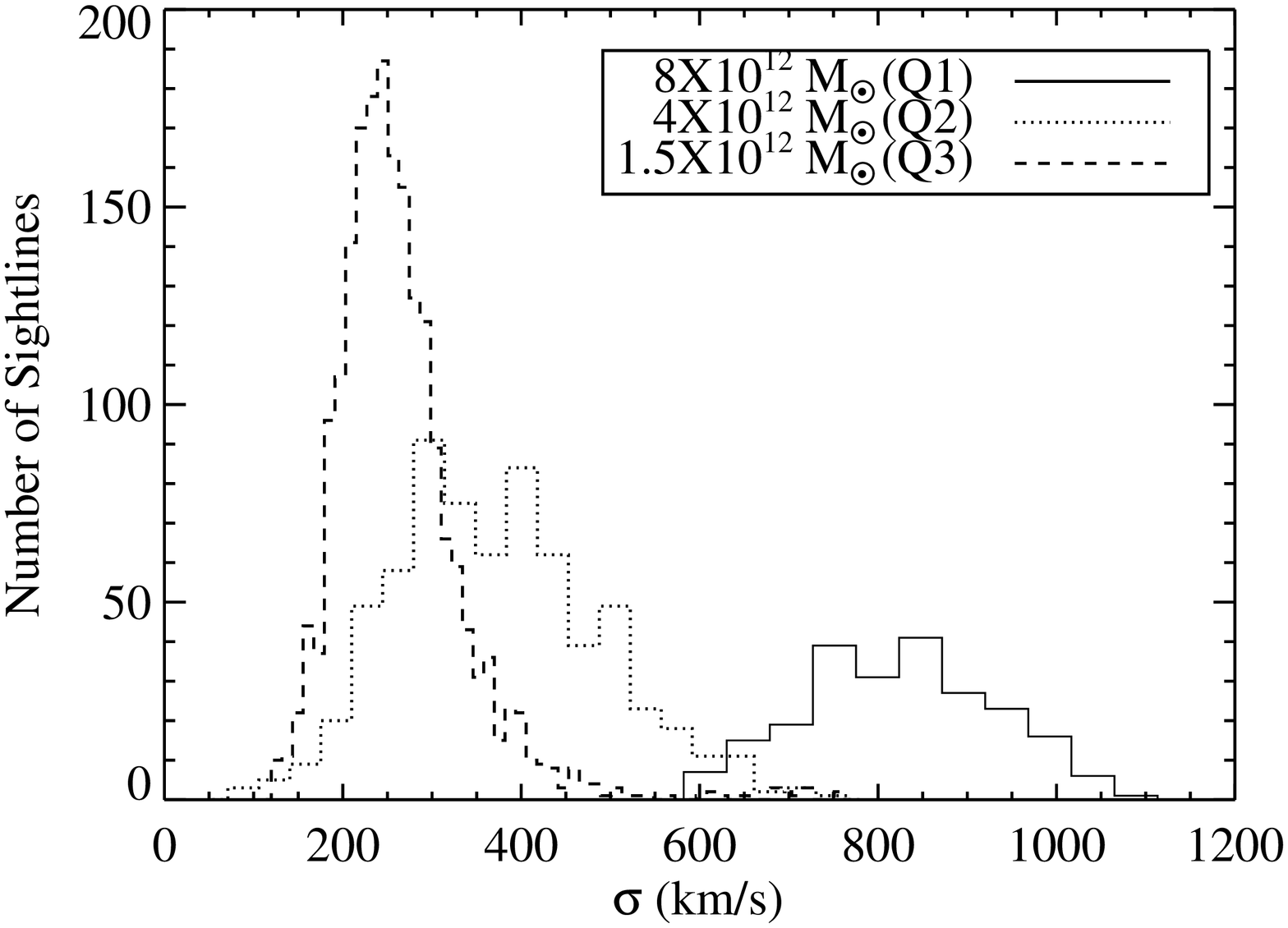}}}
\caption{Histogram of sightline-dependent line widths during quasar
phase for three halo mass models. The mean line width traces the
virial velocity of the host galaxy, and thus become narrower in lower
mass galaxies. In all cases, a broad range in linewidths is
observable. \label{figure:stdhist}}
\end{figure}

Another clear feature of Figure~\ref{figure:snap8specrandom} is the
apparent trend of narrowing line width with decreasing halo circular
velocity.  The CO linewidths are reflective of the circular velocity,
and are thus strongly dependent on the mass of the quasar host galaxy. To
further illustrate this, in Figure~\ref{figure:meanfwhm}, we show the
mean sightline-averaged CO (J=6-5) line width during the hierarchical
buildup and quasar phase of the quasars formed in the halos of Q1-Q3.
We additionally show the range of sightline-dependent observed line
width values. The multiple mergers involved in the formation of the
simulated \zsim 6 quasars give rise to large velocity dispersions
along a number of sightlines during the buildup of the quasar host
galaxies. During this time, much of the molecular gas is not
virialized, and thus the typical line widths exhibited represent about
twice the expected circular velocity at the spatial extents of the
molecular gas. In the quasar phase, as the gas virializes into a
molecular disk, the CO line widths drop to values more consistent with
the virial velocity of the host. In the most massive
($\sim 10^{13}$ \msunend, model Q1; Table~\ref{table:ICs}) halo, this
is manifested in broad ($<\sigma>\sim 500-800$ \kmsend) predicted CO
emission line widths, whereas in the halos with lower circular
velocity (quasars Q2, Q3), the mean CO line width is $\sim 450$ and
$\sim 300$ \kmsend, respectively.  It is important to note that in all
cases, while the aforementioned trends hold, a large range of line
widths is permitted at all points as they are strongly
sightline-dependent.  Another way to view this is through the detailed
distribution of line widths themselves. In
Figure~\ref{figure:stdhist}, we show a histogram of the
sightline-dependent line widths during the quasar phase for models
Q1-Q3.

These derived linewidths are a natural consequence of our initial
assumptions of quasar formation in massive halos, and that half the
cold gas mass is in molecular phase. Within the confines of our
initial assumptions, these results are consistent with virial
arguments. Thus, for example, the broad emission lines seen in the
massive $\sim 10^{13}$ \msun halo simply reflects the virial
velocity of the host galaxy at the radial extent of the molecular gas
distribution, which is of order $\sim 550$ \kmsend.

CO line measurements of \sdss by Bertoldi et al. (2003a), and Walter
et al. (2003, 2004) showed CO (J=3-2, J=6-5 and J=7-6) emission lines
with width $\sim 280 \pm 140$ \kms (FWHM; corresponding to $\sigma
\sim 120 \pm 60$ \kms for a Gaussian line). While this is a factor of
2-5 narrower than the median line width predicted in our models
(Figure~\ref{figure:meanfwhm} and ~\ref{figure:stdhist}), it is
worthwhile to note the non-negligible fraction of sightlines for each
of the quasar models that are compatible with the narrow observed line
widths (Figure~\ref{figure:stdhist}). Through the quasar phase and
beginning of post-quasar phase, $\sim2-3\%$ of sightlines in the most
massive quasar host (Q1) have linewidths compatible with the line
widths measured in \sdss (Bertoldi et al. 2003a; Walter et al. 2003,
2004).  The smaller virial velocity of the lower mass halos (Q2 and
Q3) naturally produce more sightlines with narrow line widths
compatible with observations. During the quasar phase, the quasar
formed in the $4 \times 10^{12}$ \msun halo (model Q2) shows CO
(J=6-5) emission line widths consistent with observations $\sim 5\%$ of
the time, and the quasar formed in the $1.5 \times 10^{12}$ \msun halo
(model Q3) reveals CO (J=6-5) line widths consistent with
observations $\sim 10\%$ of the time.

In an effort to understand the relationship of these models to the
measured CO line width of \sdssend, it is of interest to explore the
origin behind the particular percentages of sightlines that are
compatible with observations, and possible trends which may tend
observations toward particular sightlines.  We thus focus the
remainder of this section on this investigation. We conclude the
section with a discussion as to what circumstances may bring agreement
between our simulations and observations, and the implications of
these results with respect to potential observable tests motivated by
these models.

\subsection{Effect of Merger Remnant Structure and Disk Formation on Line Widths}
\label{section:mergerhistory}

We have thus far considered the formation of quasars hierarchically
through multiple mergers whose physical conditions were derived
self-consistently from cosmological simulations.  It is possible
that not all quasars at \zsim 6 form via numerous violent merging
events, but rather through a more 'ordered' merger. It is thus
worth quantifying the potential dependence of line width on merger
history, and in particular, how a more ordered merger remnant may
affect the observed line widths. To provide a limiting example, we
have conducted a test simulation of a coplanar binary merger in a halo
of $\sim 10^{13}$ \msun (model Q4; Table~\ref{table:ICs}). The
progenitors were initialized with a Hernquist (1990) profile, spin
parameter $\lambda$=0.033 and circular velocity $V_{\rm 200}\sim 600$
\kmsend. The disks had an initial gas fraction of 0.99, and the virial
properties were scaled to be appropriate for \z=6 (Robertson et al.
2006a).

\begin{figure}
\plotone{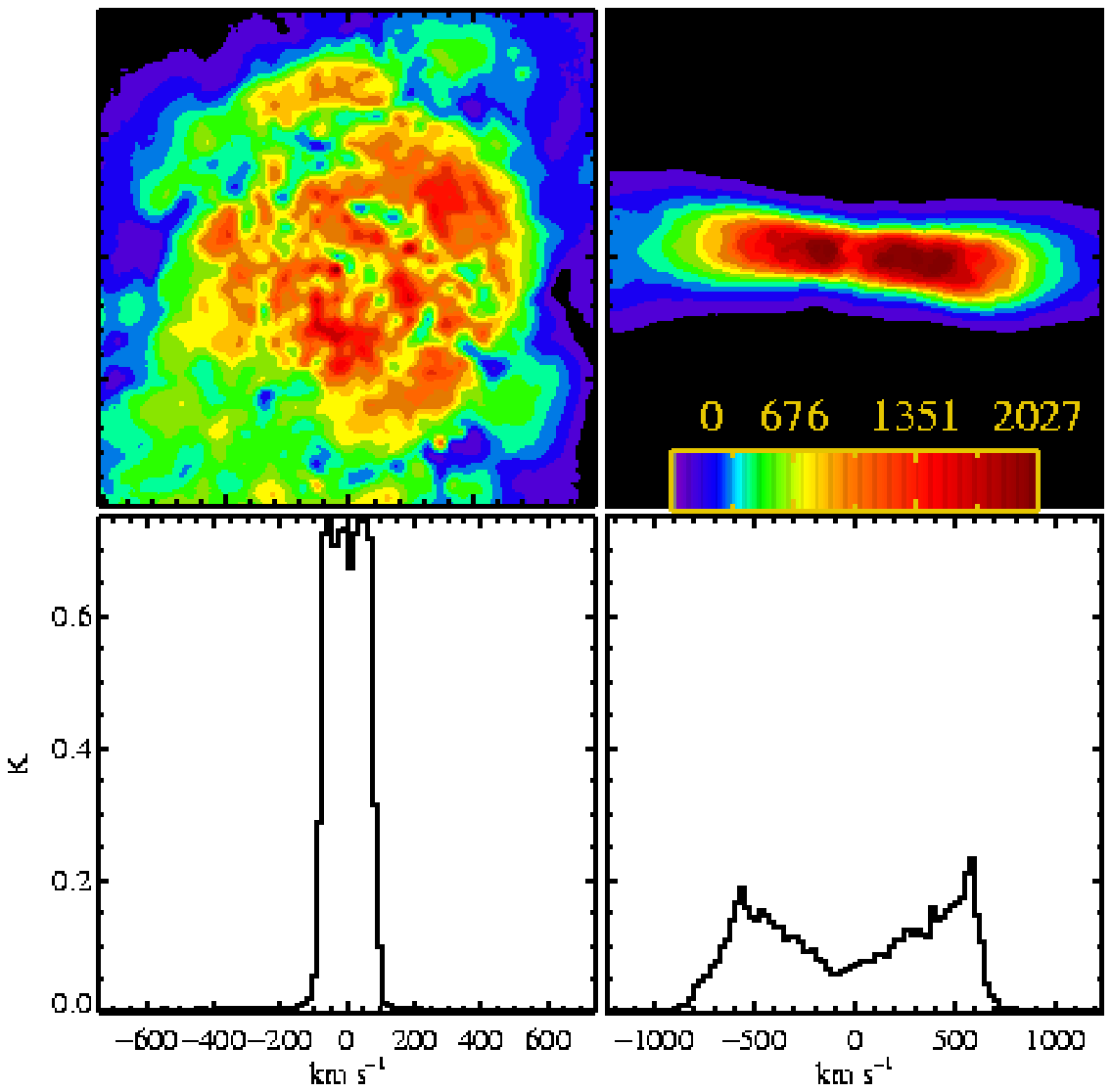}
\caption{CO (J=1-0) morphology and unresolved line profiles of remnant
formed from binary merger simulation (model Q4) across two orthogonal
sightlines (face-on and edge-on). The progenitors were initialized on
a coplanar orbit in a $\sim 10^{13}$ \msun halo.  The resultant quasar
host galaxy has large rotating molecular disk, resulting in narrow
observed emission lines in face-on viewing angles, and broad lines in
edge-on viewing angles.  Approximately 4-5\% of sightlines show narrow
line widths compatible with observations of \sdss (see
Equation~\ref{eq:probability}). The maps are 12 kpc across, and the
scale is in units of K-\kms. \label{figure:z6a6mapspec}}
\end{figure}

In Figure~\ref{figure:z6a6mapspec}, we show the CO (J=3-2) morphology
for the resultant quasar over two orthogonal viewing angles, and their
corresponding unresolved emission spectra.  The remnant forms a strong
disk-like morphology, consistent with the findings of Robertson et
al. (2006b) and Springel \& Hernquist (2005). As expected, the CO
line width and profile is a sharp function of the viewing angle of the
disk. Averaged over 250 random sightlines, we find that the predicted
CO (J=6-5) emission line width from this source is consistent with the
observed line width \sdss $\sim4\%$ of the time, comparable to the
more massive quasar hosts (Q1 and Q2) which formed from multiple
mergers.

The percentage of sightlines compatible with observations in the
binary merger Q4 arises from a limited range of angles a disk can be
from face on to keep line widths within a particular
limit. Specifically, if one considers an inclined toy disk of pure gas
with velocity dispersion $\sigma_{\rm virial}$, then in order for
observed velocity dispersions to fall below a particular value,
$\sigma_{\rm obs}$, the inclination angle from face on is limited by:
\begin{equation}
\theta < {\rm sin^{-1}}\left(\frac{\sigma_{\rm obs}}{\sigma_{\rm
virial}}\right) \, .
\end{equation}
However, inclinations along both the polar ($\theta$) and azimuthal
($\phi$) axes have to be within this limit of a face-on configuration
to keep the line of sight velocity dispersion below observed line
widths. As such, the probability of having both $\theta$ and $\phi$
randomly drawn such that they both fall below a critical value to
match an observed line width is
\begin{equation}
P(\theta <\theta_{\rm crit},\phi < \phi_{\rm crit})=\frac{4}{\pi^2}
\left({\rm sin}^{-1}\left[\frac{\sigma_{\rm obs}}{\sigma_{\rm
virial}}\right]\right)^2 \, ,
\label{eq:probability}
\end{equation}
where $P \approx4\%$ for $\sigma_{\rm obs}=180$ \kms (the upper limit
$\sigma$ of the observed line in \sdssend), and $\sigma_{\rm
virial}=600$ \kmsend. This probability then represents the upper limit
of fraction of sightlines that will be compatible with the observed CO
line width in \sdss for disks with circular velocity $\sim 500$
\kmsend.

\begin{figure}
\scalebox{.8}{\rotatebox{90}{\plotone{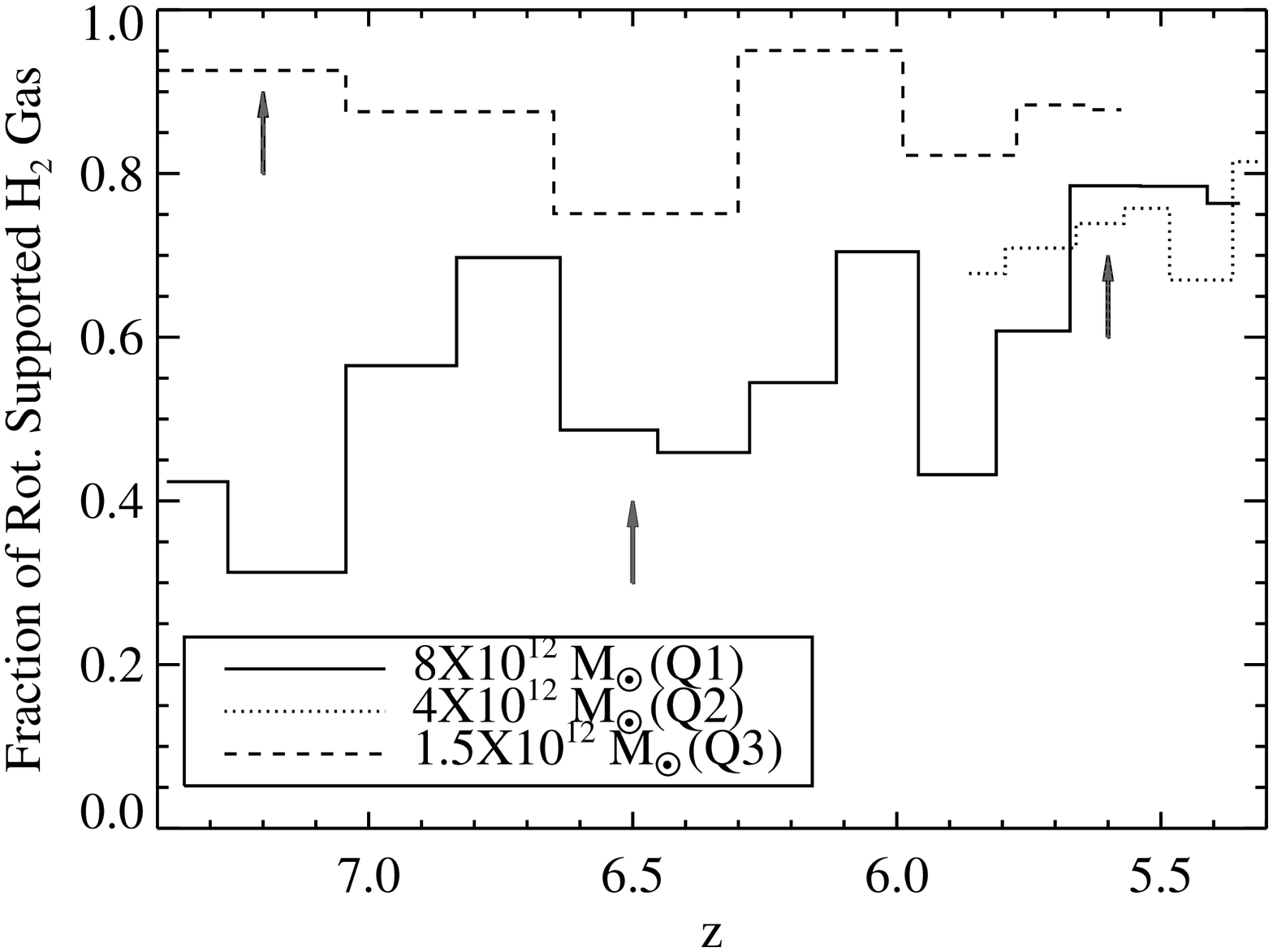}}}
\caption{Fraction of rotationally supported molecular gas in quasars
Q1-Q3. Gas is considered to be in rotational support if its rotational
velocity is at least 80\% the expected circular velocity at that
radius. Arrows point to redshift of peak quasar activity (also listed
in Table~\ref{table:ICs}). Gas is seen to more easily settle into a
stable rotational configuration in the lower mass quasar hosts, though
all quasars naturally form relatively strong disks during the quasar
phase.
\label{figure:rotsupport}}
\end{figure}

This clarifies why the percentage of sightlines compatible with
observations in the massive multiple-merger models (Q1-Q3) is
relatively small. The cold gas in the three quasars which formed out
of multiple non-idealized mergers (Q1-Q3) settles into rotating
nuclear disks. As an aside, it is interesting to note that the amount
of rotationally supported gas is seen to be dependent on the mass of
the galaxy. Namely, the lower mass halos show a larger percentage of
gas in stable rotation. To illustrate this, in
Figure~\ref{figure:rotsupport} we show the fraction of rotationally
supported gas for quasars Q1-Q3 as a function of redshift, noting in
particular the points of peak quasar activity. When the most massive
galaxy (Q1) is seen as a quasar, roughly 50\% of the \htwo \ gas is
rotationally supported. Conversely, in the lowest mass model (Q3),
$\sim 90\%$ of the gas is rotationally supported. This may be a direct
result of the amount of energy input from the central quasar during
these times. Cox et al. (2007) demonstrated that the amount of
rotationally supported gas in galaxy mergers decreases with
increasingly efficient winds. During the peak of the quasar phase, the
black hole luminosity is over an order of magnitude brighter in model
Q1 than in Q3.

In either case, though, large percentages of the gas in all of the
quasars (Q1-Q3) are seen to be rotationally supported during the quasar
phase. Because of this, the number of sightlines compatible with
observations is roughly characterized by
Equation~(\ref{eq:probability}). Thus, the most massive model (Q1) shows
line widths compatible with observations $\sim 2-3\%$ of the time,
consistent with a predicted upper limit of $\sim 4\%$ for
$\sim 10^{13}$ \msun halos. Similarly, in the lowest mass model (Q3),
where $\sim 90\%$ of the gas is rotationally supported, the 10\% of
sightlines seen to be compatible with observed line widths is
compatible with the predicted upper limit of $\sim 13\%$. The fact that
the modeled fraction of sightlines is always slightly smaller than the
toy model in Equation~(\ref{eq:probability}) owes to the fact that some
of the gas in the galaxy is still highly dynamical and not virialized.

\subsection{CO Line Width-Quasar Luminosity Relation: Potential Selection Effects}
\label{section:fwhm-qsolum}
\begin{figure}
  \scalebox{.85}{\rotatebox{90}{\plotone{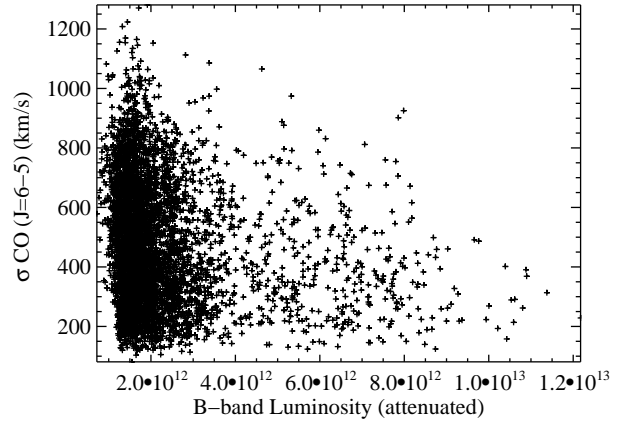}}}
  \caption{CO (J=6-5) emission line width ($\sigma$) versus attenuated
  rest-frame $B$-band luminosity. The brightest quasar luminosities
  correspond to face-on molecular disk configurations where the
  central AGN is the least obscured. These face-on sightlines
  additionally show narrow CO line widths. Thus, surveys which select
  quasars for optical luminosity may preferentially select objects
  that have narrow CO line widths. This may have important
  consequences for using CO as a dynamical mass indicator in quasars.
  \label{figure:fwhm-qsolum}}
\end{figure}

Given the dispersion of CO (J=6-5) line widths along different lines
of sight, an interesting relation to explore is one between the
optical quasar luminosity and CO emission line width in search of
potential observational selection effects which may tend observations
of \zsim 6 quasars toward smaller line widths. In
Figure~\ref{figure:fwhm-qsolum}, we plot the $\sigma$ from the CO
(J=6-5) line width as a function of the attenuated rest-frame $B$-band
luminosity over 5000 lines of sight throughout the quasar phase for
the most massive model, Q1. We utilize the methodology of Hopkins et
al. (2005a) in computing the dust-attenuated quasar luminosity, and
include contributions from both the stellar component, as well as the
central AGN.

There is a general trend for sightlines which show the brightest
rest-frame $B$-band luminosity to have smaller CO line widths. This
can be understood via decomposition of the quasar luminosity into its
stellar and AGN components. Specifically, while the stellar luminosity
does not vary much with viewing angle, the contribution of the central
black hole to the total luminosity is strongly dependent on the
viewing angle with respect to the rotating molecular gas. 
Directions which view the molecular disk in an edge-on configuration
result in a heavily obscured central AGN. Along these sightlines, the
emergent CO emission line is typically broad (e.g. bottom right panel,
Figure~\ref{figure:z6a6mapspec}). Conversely, when the molecular disk
is seen in a more face-on viewing angle (and thus has narrower CO
emission lines), the central AGN can be viewed relatively unobscured,
and the attenuated rest-frame $B$-band luminosity is consequently
higher. This effect is typical during the quasar phase, and only
rarely is the black hole relatively unattenuated through an edge-on
sightline, which causes broad CO lines to be visible when the
rest-frame $B$-band luminosity peaks.

\begin{figure}
  \scalebox{.85}{\rotatebox{90}{\plotone{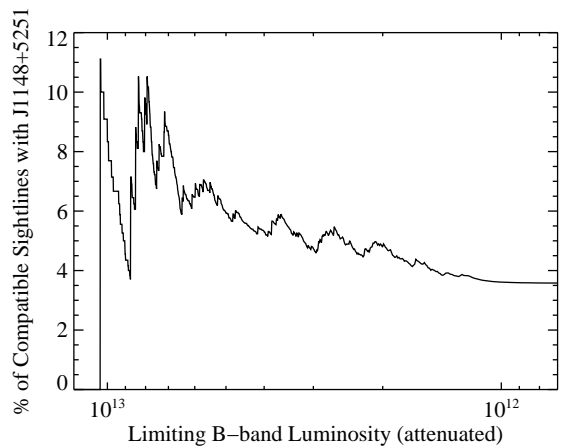}}}
  \caption{Percent of sightlines of $\sim 10^{13}$ \msun quasar over
  which line widths are compatible with observations versus limiting
  rest-frame $B$-band luminosity. Natural molecular disk formation in
  the quasar host galaxy gives rise to a selection effect which
  enhances the likelihood that flux-limited optical surveys will view
  the quasar in a face-on configuration. These viewing angles also
  typically have narrower CO line widths, and thus may increase the
  chances of viewing a narrow emission line in quasars which is not
  necessarily characteristic of the true sightline-averaged mean line
  width. Line widths are derived from CO (J=6-5) line profiles, and
  compared to upper limit of sole CO detection at \zsim 6,
  \sdssend. Rest-frame $B$-band luminosities are dust attenuated, and
  calculated using the methodology of Hopkins et al.
  (2005a). \label{figure:selection}}
\end{figure}

The results of this relationship between bolometric luminosity and
line width suggest a potential selection effect which may cause
quasars selected for optical luminosity to have systematically lower
CO line widths, owing to the preferred face-on viewing angle for the
molecular disk. In Figure~\ref{figure:selection}, we plot the
percentage of sightlines with line widths compatible with observations
as a function of limiting rest-frame $B$-band luminosity. For the
highest flux cuts, the fraction of sightlines with narrow line widths
increases from 2-3\% to $\sim10\%$. This selection effect is robust
across both the lower mass halo models as well. In models Q2 and Q3,
the fraction of sightlines compatible with observations increases to
$\sim 15$ and 25\% respectively, for the highest luminosity cuts.

Finally, we note that the viewing angles corresponding to the smallest
($2\sigma$) line widths in our simulations typically fall within a
$25\degr$ range in polar and azimuthal angle from face-on. This range
of angles is consistent with recent observational studies which have
suggested that the commonly observed narrow CO line widths of high-\z
\ quasars may correspond to a preferred viewing angle of $10-15\degr$
from face-on (Carilli \& Wang 2006; Wu 2007).

\subsection{Interpretation of \zsim 6 Quasar Observations}
\label{section:interpretation}

Our simulations show that \zsim 6 quasars which form in massive halos
will characteristically have broad mean line widths, consistent with
simple virial arguments. The mean simulated line widths are in
apparent contradiction to observed narrow ($\sigma \sim 120 \pm 60$
\kmsend) line widths of the sole CO detection at \zsim 6 (Walter et
al. 2004), though a non-negligible fraction of sightlines (ranging
from 2-3\% for the most massive model to $\sim10\%$ in the lowest mass
model) are compatible with this observation. Potential selection
effects owing to molecular disk formation in these galaxies will
increase the probability of narrow-line detection when selecting
quasars for optical luminosity, ranging from $\sim 10\%$ in the most
massive quasar host to 25\% in less massive ones. The full dispersion
in line widths is predicted to become more apparent at lower optical
luminosities (Figure~\ref{figure:fwhm-qsolum}).

Our models find that quasars which form in the lower end of our halo
mass range at \zsim 6 (e.g. quasars Q2 and Q3) may have similar rates
of detection as those which form in the most massive
$\sim 10^{13}$ \msun halos (e.g. quasar Q1). This owes to the
competing effects of smaller quasar lifetimes in the lower mass halos,
but more lower mass halos in the simulated cosmological
volume. Specifically, the quasars formed in the
$4 \times 10^{12}$($1.5 \times 10^{12}$) \msun halos have luminosities
$\ga 10^{13}$ \lsun for a factor of $\sim 2.5(14)$ less time than the
quasar formed in the $10^{13}$ \msun halo
(Table~\ref{table:ICs}). However, standard halo mass functions predict
more low mass halos than massive $\sim 10^{13}$ \msun halos (Press \&
Schechter 1974; Sheth \& Tormen 2002; Springel et al. 2005). The
cosmological simulations of Li et al. (2007) (which were used in this
work) found approximately 3(7) times as many halos of mass
$5 \times 10^{12}$($2 \times 10^{12}$) \msunend, compared to the single
$\sim 10^{13}$ \msun halo identified in the simulation box at \zsim 6
(Li et al. 2007, Figure 14). In this sense, quasars which form in
lower mass halos may be identified at similar rates as those
which form in $\sim 10^{13}$ \msun halos.

One result of this work is to motivate observational tests of these
models. A direct prediction of these simulations is that a CO survey
from a large sample of quasars at \zsim 6 which probes lower on the
optical luminosity function may directly constrain the range of
potential line widths which originate from high redshift quasars. At
the median rest-frame $B$-band luminosity, our models predict that a
large range of line widths should be observed
(Figure~\ref{figure:fwhm-qsolum}). A potential caution associated with
this test is that physical processes on scales below the resolution of
our simulations may limit optical detections of quasars with edge-on
disks at \zsim 6. For example, if a dusty molecular torus exists on
scales smaller than $\sim 100$ pc and provides high levels of
obscuration along sightlines other than face-on, then the variation of
optical luminosity with inclination angle will be steeper than
suggested by Figure~\ref{figure:fwhm-qsolum}. It is thus not a
straightforward assumption that optical surveys will need to probe
only an order of magnitude lower in rest-frame $B$-band luminosity to
test these models as Figure~\ref{figure:fwhm-qsolum} suggests, but
rather it is in the limit that quasars with relatively inclined disks
at \zsim 6 can be detected that these models predict a broad range of
CO line widths at lower optical luminosities.

A more clear test may come from observations of the progenitors of
\zsim 6 quasars themselves. For example, a direct prediction from
these models is that CO observations of either the most massive
progenitor galaxies prior to the merger (e.g. at \zsim 8; see Table 1
of Li et al. 2007), or the ongoing mergers themselves may exhibit a
large dispersion in CO line widths, with median velocity dispersion
reflective of the halo virial velocity
(Figure~\ref{figure:meanfwhm}). The identification of potential
progenitors at \zga 7 is predicted to be feasible through $z$-band
dropouts (Robertson et al. 2007).


In either case, these models suggest that surveys at \zsim 6 which
observe either the most massive progenitors of quasar host galaxies, or
highly inclined disks (closer to edge-on, and likely lower luminosity)
associated with \zsim 6 quasars will see a broader dispersion of CO
line widths. The model halos presented here reflect the range of halo
masses in our cosmological simulation which were feasibly able to
create a \zsim 6 quasar. As such, based on the line widths seen in our
lowest mass host galaxy (Q3), samples of CO detections at \zsim 6
which probe quasars with a range of disk inclination angles should
find a median line width at least $\sim$ twice the value of the sole
detection. Surveys which observe quasars with inclined molecular disks
and still find consistently narrow line widths may reflect an
inability of our radiative transfer simulations to fully capture the
appropriate physics necessary to predict accurate CO line widths form
these early Universe galaxies. If, for example, our assumptions
regarding molecular gas content or CO abundances are incorrect, it
could be that CO emission is preferentially seen in lower velocity gas
in the host galaxy. Alternative possibilities may include
super-Eddington accretion for the central black hole which may allow
for luminous quasars at \zsim 6 in less massive halos than those adopted
here (e.g. Volonteri \& Rees 2005). It is, however, attractive that
the simulations presented here do provide a model for the CO emission
from the earliest quasars which are consistent with observations of
the CO excitation and morphology of \sdssend. More observations to
fully determine the nature of CO line widths in \zsim 6 quasars will
be necessary to assess the validity of this aspect of our
modeling.

\section{Spheroid, Black Hole, and Dynamical Mass}
\label{section:mdyn}
Because submillimeter-wave radiation typically does not suffer the
heavy extinction characteristic of optical emission, CO lines are
often used as dynamical mass indicators in dusty starburst
galaxies. Dynamical masses are derived assuming that the emitting gas is
rotationally supported, and that the line width provides a measure of
the rotational velocity. Here we assess the usage of CO-derived
dynamical masses in \zsim 6 quasars. As a case study, we will focus on
the most massive simulation ($\sim 10^{13}$ \msunend; model Q1),
though the trends are generic for all quasars presented in this work.

Generally, theoretical arguments have predicted that the $M_{\rm
BH}$-$\sigma_{\rm v}$ and \magorrian relations show only weak
evolution with redshift (Robertson et al. 2006a). Using numerical
simulations of galaxy mergers by Cox et al. (2006b) and Robertson et
al. (2006a) that include black hole feedback, Hopkins et al. (2007a)
found that the normalization of the $M_{\rm BH}$-$M_{\rm bulge}$
relation shows weak ($\sim 0.3-0.5$ dex) trends toward larger $M_{\rm
BH}$/$M_{\rm bulge}$ from \z = 0 - 6. The simulations presented here
provide additional support for a scenario in which the stellar bulge
and central black hole grow coevally in the earliest galaxies.

In the most massive galaxy presented here, the supermassive black hole
grows rapidly during the hierarchical buildup of the quasar host
galaxy, and reaches a total mass of $\sim 2 \times 10^{9}$ \msun
during the peak quasar phase, similar to black hole mass estimates in
\sdss (Willott et al. 2003). Owing to the extreme star formation
rates which can be as large as $\sim 10^4$ \msunyr between redshifts
\z=9 and \z=8 during the final violent mergers (e.g,
Figure~\ref{figure:lco}), the bulge reaches a total mass of
$\sim 10^{12}$ \msun by the quasar phase. The black hole and stellar
bulge masses are related such that $M_{\rm BH}=0.002\times M_{\rm
bulge}$ during the quasar phase, roughly consistent with the local
\magorrian relation (Li et al 2007; Magorrian et al. 1998; Marconi
\& Hunt 2003).  We note, though, that during the peak of the quasar
activity, the $M_{\rm BH}-\sigma_{\rm v}$ relation is not
necessarily obeyed as the stellar bulge is not dynamically relaxed.

 The massive starburst at \zga 7 results in an ISM highly enriched
with metals, consistent with the observed FeII/MgII abundances, [CII]
emission CO emission, and dust masses for \zsim 6 quasars (Barth et
al. 2003; Bertoldi et al. 2003b; Dietrich et al. 2003a;, Freudling
et al. 2003; Jiang et al. 2006; Maiolino et al. 2005; Walter et
al. 2003, 2004). Li et al. (2007) have found that the metallicity in
the simulated $\sim 10^{13}$ \msun quasar presented here is solar to
supersolar during the quasar phase, owing to the $\sim 10^4$ \msunyr
starburst during the hierarchical merging process of the quasar
formation. These findings are consistent with the mean abundances of
$\sim 4$ times solar in a sample of $4<$\z$<5$ quasars observed by
Dietrich et al. (2003b) and imply that black hole growth and stellar
bulge formation are correlated at high redshifts. These simulations
support findings that the relationship is regulated by feedback from
supermassive black holes (e.g. Di Matteo, Springel \& Hernquist 2005;
Robertson et al. 2006a; Hopkins et al. 2007a).

CO observations of high-\z \ quasars have suggested that central black
hole masses may be excessively large compared to the stellar bulge
mass as predicted by the present day \magorrian relation. Dynamical
mass estimates of \sdss using CO line measurements have indicated that
the stellar bulge may be undermassive by a factor of $\sim 10-50$ if
the present day \magorrian relation holds at \zsim 6 (Walter et al.
2004). Studies of other quasars at \zga 3 using CO line widths as a
proxy for enclosed mass have arrived at similar conclusions (Shields
et al. 2006). 

\begin{figure}
\scalebox{1.2}{\plotone{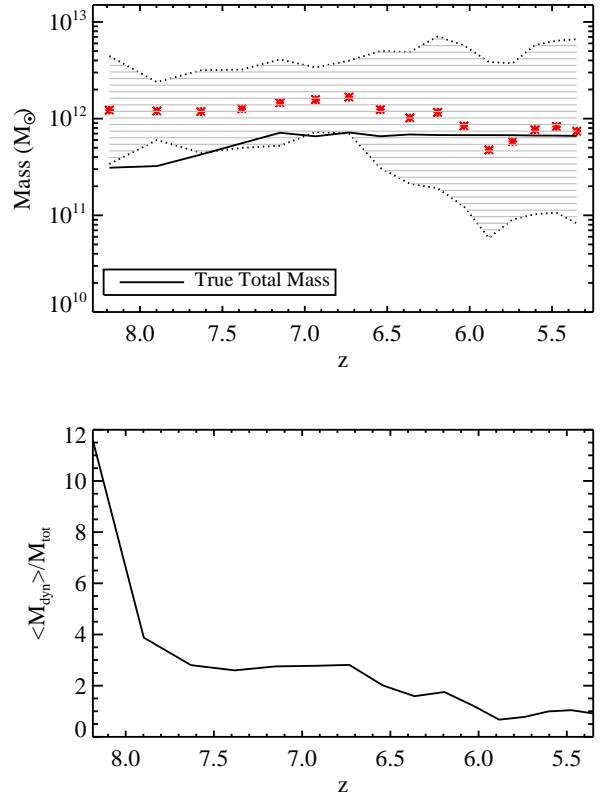}}
\caption{{\it Top:} Median dynamical masses derived from line widths
over 250 random sightlines through the hierarchical buildup and
evolution of quasar host galaxy (red crosses). The shaded region
represents the $3\sigma$ range of sightline-dependent derived
dynamical masses, and the solid line the true total mass within the
central 2 kpc. The dynamical masses are derived assuming an
inclination angle of $30\degr$, and a molecular spatial extent of 2
kpc. {\it Bottom:} Ratio of median derived dynamical mass to total
mass. During the hierarchical buildup and early quasar phase, much of
the gas is highly dynamical; consequently, derived dynamical masses
from CO line widths do a poor job representing the true enclosed mass,
typically overestimating by factors of 2-5. As the gas virializes the
sightline-averaged CO line width is a better estimator of the
dynamical mass. A large range of values is permitted, consistent with
the spread in CO line widths seen in Figure~\ref{figure:meanfwhm}.
\label{figure:mdynhist}}
\end{figure}

In order to investigate the usage of CO as a dynamical mass indicator,
in Figure~\ref{figure:mdynhist}, we plot the evolution of the median
dynamical mass of the quasar host galaxy from the most massive
simulation (Q1) derived from the CO (J=6-5) emission line over 250
random lines of sight, as well as $3\sigma$ contours for the observed
range of linewidths. In particular, we show
\begin{equation}
M_{dyn}=k\frac{\sigma^2 R}{G \ {\rm sin}(i)}
\end{equation}
where we use an adopted $k$ of 8/3 and assume an average inclination
angle of $30 \degr$. We plot the total mass of the host galaxy (within
the central 2 kpc) throughout its evolution, including the multi-phase
ISM, dark matter, stars, and black holes. We additionally plot the
ratio of the median derived dynamical mass to the total mass
enclosed. The dynamical mass is derived from the width of the CO
(J=6-5) line as the CO excitation peaks at the J=6 level, and this
transition thus best trace the bulk of the molecular gas during the
quasar phase of our simulated galaxy. We note, however, that the line
widths from lower transition lines do not deviate much from these
trends.

During the hierarchical buildup of the host galaxy (pre-quasar phase),
the gas is still highly dynamical, and not completely virialized. This
results in large CO line widths, and consequently derived dynamical
masses which largely overestimate the true mass. During this time, the
typical derived dynamical mass results in an overestimate of the true
mass by a factor of $\sim 2-5$. As the gas becomes more rotationally
supported during the quasar phase, the CO (J=6-5) line width serves as
a better tracer of the mass enclosed. These results are consistent
with similar derivations by Greve \& Sommer-Larsen (2006) who found
that CO can serve as an accurate dynamical mass tracer in merger
simulations to within 20\%.

An important point from Figure~\ref{figure:mdynhist} is that a large
sightline-dependent range of values are possible for derived dynamical
masses. While the median line widths provide reasonable estimates of
the true mass, many lines of sight permit significant underestimates.
The radiative transfer simulations presented here may then bring some
resolution between CO observations which are suggestive of a strong
evolution in the \magorrian relation at high redshift, and models
which imply a lesser evolution between present epochs and early times.
Figure~\ref{figure:mdynhist} shows that a large range of dynamical
masses may be inferred simply based on observed viewing angle. While
the median CO line widths accurately traces the enclosed mass for the
bulk of the quasar phase, potential selection effects
(Figures~\ref{figure:fwhm-qsolum} and ~\ref{figure:selection}) may
bias masses derived from CO line widths from quasars toward an
underestimate of the true mass. These model results are consistent
with the recent study by Wu (2007) who found that CO line widths
provided a particularly poor estimate of bulge velocity dispersion for
local Seyferts with line widths narrower than $\sim 400$ \kms (FWHM;
$\sigma \sim 170$ \kmsend). In contrast, the inclination-corrected CO
line width was found to correspond well with the bulge velocity
dispersion.


\section{Comparisons to Other High Redshift Populations}
\label{section:discussion}

A natural question which arises from this work is, how do the CO
properties of these extreme \zsim 6 objects relate to other known
starburst and AGN populations? Two extreme high redshift galaxy
populations which may serve as interesting comparisons are:
1. Optically selected quasars; and 2. Dusty submillimeter selected
galaxies at \zsim 2 (see respective reviews by Solomon \& Vanden Bout,
2005, and Blain et al. 2002).

\subsection{Quasars}
Only a handful of quasars have been detected in molecular line
emission owing to beam dilution and limited sensitivity at millimeter
and submillimeter wavelengths. As of the writing of the recent review
by Solomon \& Vanden Bout (2005), there have been about 16 quasars at
\zga 1 for which molecular gas emission properties have been published.

Deciphering the molecular gas morphology remains an issue for most
quasars since the majority of them have been detected in molecular
line emission with the help of gravitational lensing, resulting in
multiple imaging. The spatial extent of the molecular emission in most
quasars appears to range from $\sim 1-5$ kpc (Solomon \& Vanden Bout,
2005), similar to the simulations presented here as well as observations
of \sdssend. Multiple CO emission peaks are detected less often,
though it is not clear whether or not this is a spatial resolution
issue. Of seven imaged quasars listed in the recent review by Solomon
\& Vanden Bout (2005), three have clear CO companions (including
\sdssend), with distances in the emission peaks ranging from 1.7-8.7
kpc (e.g. Carilli et al. 2002).  The more sensitive, and higher
spatial resolution observations that will be routinely achieved with
ALMA will clarify the CO morphology of quasars.

The CO excitation conditions that have been measured in quasars have
primarily been at \zsim 2. Barvainis et al. (1997) find that the CO is
substantially excited in the Cloverleaf quasar (H1413+117, \zsim
2.5) through the J=7 level. High spatial resolution observations of
two \zsim 4 quasars by Carilli et al. (2002) find similar
excitation conditions in CO, both of which have multiple CO
emission peaks in their morphology. More recent large velocity gradient
modeling of \zsim 4.7 quasar BR 1202-0725 also found a CO SED that
peaks at J=7 (Riechers et al. 2006b). The high excitation conditions
observed in quasars are indicative of extremely warm and dense
molecular gas heated by ongoing star formation, and are consistent
with active star formation during the quasar phase.

The quasar sample of Solomon \& Vanden Bout reports CO line widths
in high-\z \ quasars ranging from roughly $\sigma \sim 100-250$ \kms
(Carilli \& Wang 2006). Large bandwidth observations with the 4GHz
COBRA correlator on OVRO by Hainline et al. (2004) of a sample of
three \z=2-3 quasars showed similarly narrow CO line width.

While these are in apparent contrast to the large mean line widths in
\zsim 6 quasars reported here, the narrower line widths of quasars
from $2\leq z \leq5$ may be simply explained by the evolution of $V_c$
with decreasing redshift. Namely, while quasars from $2\leq z \leq5$
appear to form in halos of comparable mass to those at \zsim 6 (Croom
et al. 2005; Shen et al. 2007), the expected circular velocity will
naturally be lower than their \zsim 6 analogs by a factor of
$\sqrt{(1+z)/(1+6)}$. As a result, the circular velocity for these
halos at \zsim 2 would be of order $\sigma \sim 300-350$ \kmsend,
consistent with clustering measurements of \zsim 2 quasars (e.g. Croom
et al. 2005; Porciani \& Norberg 2006; Myers et al. 2006;
Hopkins et al. 2007d) and
studies of the quasar proximity effect (e.g. Faucher-Giguere  et al. 2007;
Kim \& Croft 2007; Nascimento Guimaraes et al. 2007) ,
and
slightly larger than measured CO line widths (Solomon \& Vanden Bout,
2005), though it is possible that there is a potential selection bias
toward quasars with narrow line widths (\S~\ref{section:fwhm-qsolum})
at lower redshifts as well.

In this context, we emphasize that the simulations presented here {\it
do not} predict large ($\sigma \sim 500-800$ \kmsend) line widths for
quasars which form at lower (e.g. \zsim 2) redshift, even if they form
in halos of comparable mass to luminous sources at \zsim 6. To
illustrate this, we examine the predicted CO line widths of the binary
merger simulation of Narayanan et al. (2006a; also presented in
Robertson et al. 2006a). The circular velocity of the progenitor
galaxies was $V_{\rm c}$=160 \kmsend. The total masses of the
progenitor galaxies in this example was $4.8 \times 10^{11}$
 \hmsunend, and the final merger produced a central black hole mass of
$\sim 5 \times 10^7$ \hmsunend. The $\sigma$ of the unresolved CO
(J=1-0) line profile ranges from 100-150 \kms during the peak of the
quasar phase, over three orthogonal viewing angles, consistent with CO
line width measurements of \zsim 2 quasars, and reflective of the host
virial velocity.

These models then suggest a self-consistent picture which naturally
explains the evolution of CO line widths as function of redshift in
terms of the circular velocity of the quasar host galaxy halo. Future
modeling will need to quantify the extent of potential molecular
disk-driven selection effects (e.g. \S~\ref{section:fwhm-qsolum}) for
quasars at lower redshifts (\zsim 2-4).  Indeed other works have
suggested that quasars in this redshift range may be observed at a
preferred shallow range of viewing angles as evidenced by their CO
line widths (e.g. Carilli \& Wang 2006).

\subsection{Submillimeter Galaxies}
Submillimeter galaxies (SMGs) represent a class of massive, dusty
starbursts at \zsim 2 largely detected by blind surveys with the SCUBA
and MAMBO bolometers. Typical infrared luminosities in these sources
of $\sim 10^{13}$ \lsun correspond to a SFR of $\ga1000$ \msunyrend,
assuming an insignificant AGN contribution (Smail 2006). X-ray and IR
studies have shown that these galaxies are known to contain embedded
AGN, although their relative contribution to the bolometric luminosity
is uncertain (Alexander et al. 2005a,b; Donley et al. 2005; Polletta
et al. 2006).  Optical morphologies of SMGs indicate that many are
interacting and/or mergers (Chapman et al. 2004). Recent CO
morphologies and emission line profiles have furthered this scenario
(Greve et al. 2005; Tacconi et al. 2006). Many studies have pointed
to a picture in which SMGs may be ongoing mergers at \zsim 2, though
in a pre-quasar phase (e.g. Blain et al. 2002). This combined with
the fact that SMGs are the most massive and actively star forming
galaxy population at \zsim 2 make SMGs an interesting comparative for
our simulated \zsim 6 quasar.

The excitation characteristics have been observed in multiple CO
emission lines for only one case: \z=2.5 SMG SMM J16359+6612. In this
galaxy, the CO flux density peaks at J=5 (Wei\ss \ et al. 2005a),
consistent with highly excited gas. The excitation conditions are
similar to those seen in our simulations when the SFRs are comparable.

The molecular morphology in SMGs closely resembles the extended CO
emission seen in simulations and observations of \sdssend, as well as
in the pre-quasar phase galaxy in our simulations at \z=7-10. The
average CO FWHM radius in the recent Tacconi et al. (2006) sample is 2
kpc.  The large spatial extent of CO emission in SMGs combined with an
apparent lack of strong quasar activity has been interpreted as being
the consequence of extremely massive and gas rich galaxy mergers early
in the evolution of the galaxy, and prior to the quasar phase
(e.g. Tacconi et al. 2006). Indeed, if SMGs form through hierarchical
mergers, then the models presented here suggest that the spatially
extended and disturbed CO morphologies seen in SMGs further indicate
that these galaxies may be mergers prior to an optical quasar phase
(e.g. Figure~\ref{figure:mapevolution}).

The CO line widths of SMGs are more enigmatic. The average CO emission
line from an SMG is typically about double the width of that from a
quasar of similar redshifts (e.g. Carilli \& Wang 2006; Greve et al.
2005; Tacconi et al. 2006).  The average line width in SMGs is
$\sigma \sim 330$ \kms (${\rm FWHM}\sim 780$ \kmsend), compared to a mean
$\sigma$ of 130 \kms seen in \zsim 2 quasars (Greve et al.
2005). Within the context of the models presented in this work, a
number of physical motivations for these differing line widths may be
at play.

First, in our models the mean line widths from galaxies with
virialized cold gas are seen to roughly correspond with the circular
velocity of a host halo. When comparing galaxies of similar redshifts,
if SMGs are dynamically relaxed, this would imply that SMGs may
originate in more massive halos than quasars. Indeed, clustering
measurements made by Blain et al. (2004) have suggested that SMGs are
hosted by halos slightly larger than quasars at comparable
redshifts. If SMGs reside in halos $\ga$4 times more massive than
typical quasar host galaxies, the difference in CO line widths may
be accounted for.

An alternative explanation for the line widths may arise from an
evolutionary standpoint. Specifically, as Figure~\ref{figure:meanfwhm}
shows, there is a typical drop in CO line width by a factor of $\sim 2$
close to the quasar phase of a merging galaxy system. This owes to gas
in the violent environs of a galaxy merger becoming rotationally
supported late in the evolution of the merger. If SMGs are a class of
objects hosted by halos of similar mass to their quasar counterparts,
then it may be that SMGs are simply massive mergers at \zsim 2 prior
to their active quasar phase. Certainly CO line profiles and
morphologies from SMGs are consistent with merging activity (Greve et
al. 2005; Tacconi et al. 2006; Narayanan et al. 2006; D. Narayanan
et al., in prep). Moreover, numerical simulations of merging
galaxies coupled with self-consistent radiative transfer solutions
have pointed to a picture in which SMGs may be mergers caught in a
phase of massive black hole growth, though prior to an optical quasar
phase (Chakrabarti et al. 2007b).

Other authors have suggested that SMGs may be similar to quasars in
their place in galaxy evolution, but simply viewed more edge-on.
Carilli \& Wang (2004) suggested that if these \zsim 2
galaxy populations are truly of the same class, then one can infer a
mean viewing angle of $\sim 13 \degr$ for quasars, whereas SMGs are
more likely randomly oriented.

Finally, it may be that the CO emission properties of SMGs are not
explained by physical models such at those presented in this work. If,
for example, the massive star formation rates are fueled at least in
part by accretion of gas from the host halo (which is seen to occur
for at least some galaxies in cosmological simulations; see
e.g. Finlator et al. 2006), the CO line widths may not reflect the
circular velocity of the system, at least during phases of elevated
star formation, as they do in merging galaxies.

In summary, correspondence between some observed trends of molecular
line emission in the models of hierarchical \zsim 6 quasar formation
and SMGs which form at later times suggest that it is plausible that
SMGs fall naturally into a merger-driven evolutionary sequence,
though, at least from molecular line diagnostics alone, its location
on this sequence is not completely clear. Of course, alternative
scenarios cannot be ruled out here. Further models of hierarchical galaxy
mergers appropriate for \zsim 2 will have to be examined in order to
further quantify the relation of CO emission properties to the
evolution of SMGs.

\section{Summary and Conclusions}
\label{section:conclusions}

We have applied non-LTE radiative transfer simulations to cosmological
and hydrodynamic galaxy formation simulations to predict the CO
emission from representative \zsim 6 quasars that are modeled to form
hierarchically in massive $10^{12} - 10^{13}$ \msun halos. We made
predictions concerning the CO excitation patterns, morphologies, and
line widths in this extreme class of objects. We further made broad
comparisons to the only current CO detection at \zga 6, \sdss at
\z=6.42. Our main results are the following:

\begin{enumerate}

\item Owing to very warm and dense conditions in the molecular ISM,
the CO flux density is predicted to peak at the J=8 level during the
early, hierarchical formation process of the quasar host galaxy (\zsim
8), when the SFR can be as high as $\sim 10^4$ \msunyrend. During the
peak quasar phase, the central AGN reduces the nuclear starburst, and
the SFR drops to $\sim 10^2$ \msunyrend. Consequently, the CO flux
density peaks at the J=5-6 level. These excitation conditions are
indicative of an ongoing starburst, and are consistent with
observations of \sdssend. As the gas becomes more diffuse and the
starburst dies down in the post-quasar phase (\zla 6), the peak in the
CO flux is predicted to drop to J$\approx 3$.

\item The CO morphologies of \zsim 6 quasars may exhibit multiple
emission peaks during the active quasar phase, owing to separated
peaks of high density emission that have not yet coalesced.  The
multiple emission peaks in the morphology of the CO (J=3-2) gas during
the quasar phase of our simulations are very similar to observations
of \sdssend, and are robust along many viewing angles.  These results
imply that a merger-driven formation scenario for \zsim 6 quasars
produces CO morphologies consistent with that of \sdssend.

\item On average, the CO line widths from quasars at \zsim 6 are
reflective of the virial velocity of the host halo, though there
exists a large sightline-dependent dispersion in line widths. During
the hierarchical buildup of the host galaxy, the median line widths
are roughly twice the virial velocity, and settle to the virial
velocity during the quasar phase. During the quasar phase, the
sightline-averaged line width for the $\sim 10^{13}$ \msun halo is
$\sim 500-800$ \kmsend. In the lowest mass halo ($\sim 10^{12}$
 \msunend) the sightline averaged line width is $\sim 300$ \kmsend.

\item A fraction of sightlines in each model is compatible with
 observations, and is a strong function of halo mass. Specifically,
 the number of sightlines with narrow line widths compatible with
 observations increases with decreasing halo mass. The most massive
 $\sim 10^{13}$ \msun halo shows $\sim 2-3\%$ of sightlines compatible
 with observations. The $\sim 10^{12}$ \msun halo has line widths
 similar to observations of \sdss $\sim 10\%$ of the time. The
 percentage of sightlines compatible with observations increase owing
 to selection effects (next point).

\item Quasars at \zsim 6 selected for optical luminosity may
preferentially be in a face-on configuration as this provides the
least obscuration of the central black hole. In these configurations,
the CO line widths are narrower, thus causing quasars selected for
optical luminosity to preferentially have narrower line widths than
their sightline averaged values. The fraction of sightlines with line
widths compatible with observations increases to 10-25\% when
considering quasars selected for optical luminosity. This suggests
that these models may be in agreement with observations if \sdss is
being observed in a face-on configuration.  A direct consequence of
these selection effects is that in order to observe the full
dispersion in CO line widths in \zsim 6 quasars, observations must
probe quasars with edge-on molecular disks (which are typically lower
on the optical luminosity function).

\item Because of the evolution of $V_c$ with redshift of halos of
  similar mass, quasars which form in $\sim 10^{12} - 10^{13}$ \msun
  halos (as they are thought to; see e.g. Croom et al. 2005; Shen et
  al. 2007) at lower redshift will naturally have smaller line
  widths, compatible with observations. We explicitly show this by
  examining the line widths of a binary merger simulation appropriate
  for present epochs. In this light, the simulations presented here
  {\it do not} predict large ($\sigma \sim500-800$ \kmsend) CO line
  widths for quasars which form at lower redshifts, but rather a
  suggestive self-consistent model for the potential origin of CO line
  widths in quasars at both low and high redshift.

\item Our merger-driven model for quasar formation predicts a host
galaxy that lies on the \magorrian relation during the active quasar
phase (Li et al. 2007; Robertson et al 2006a; Hopkins et al.
2007a). During the hierarchical buildup of the host galaxy, the median
CO line width tends to typically overestimate the dynamical mass by a
factor of 2-5 as much of the gas is highly dynamical, and not
virialized. During the quasar phase, dynamical masses derived from the
median line widths are a better representation of the true mass. There
is a large range in derived dynamical masses coincident with the large
sightline-dependent range of line widths seen at a given time. If
selection effects are in place such that molecular disks in observed
high-\z \ quasars are typically close to face-on, CO-derived dynamical
masses will preferentially underestimate the true mass unless the
shallow viewing angle is accounted for.

\end{enumerate}

\acknowledgements DN would like to express appreciation to Chris
Carilli, Xiaohui Fan, Reinhard Genzel, Brandon Kelly, Yancy Shirley,
Andy Skemer, Volker Springel and Fabian Walter for helpful
conversations. DN was supported for this study by an NSF graduate
research fellowship. BER gratefully acknowledges the support of a
Spitzer Fellowship through a NASA grant administered by the Spitzer
Science Center.  This work was supported in part by NSF grant AST
03-07690 and NASA ATP grant NAG5-13381. Support for this work was also
provided by NASA through grant number HST-AR-10308 from the Space
Telescope Science Institute, which is operated by AURA, Inc. under
NASA contract NAS5-26555. The calculations were performed in part at
the Harvard-Smithsonian Center for Parallel Astrophysical Computing.

\end{document}